\documentclass{article}

\usepackage{microtype}
\usepackage{graphicx}
\usepackage{subcaption} % provides the subfigure environment
\usepackage{booktabs} % for professional tables
\usepackage{hyperref}

\usepackage[accepted]{icml2025}

\usepackage{amsmath}
\usepackage{bm}
\usepackage{amssymb}
\usepackage{mathtools}
\usepackage{amsthm}
\usepackage{enumitem}
\usepackage[capitalize,noabbrev]{cleveref}

\theoremstyle{plain}

\theoremstyle{definition}

\theoremstyle{remark}

\usepackage[textsize=tiny]{todonotes}

\begin{document}

\twocolumn[
\icmltitle{Continual Learning for Wireless Channel Prediction}

\icmlsetsymbol{equal}{*}

\begin{icmlauthorlist}
\icmlauthor{Muhammad Ahmed Mohsin}{equal,stan}
\icmlauthor{Muhammad Umer}{equal,stan}
\icmlauthor{Ahsan Bilal}{equal,ou}
\icmlauthor{Muhammad Ali Jameshed}{sch}
\icmlauthor{John M. Cioffi}{stan}
\end{icmlauthorlist}

\icmlaffiliation{stan}{Department of Electrical Engineering, Stanford University, Stanford, CA 94305, USA}
\icmlaffiliation{ou}{School of Computer Science, University of Oklahoma, Norman, OK 73019, USA}
\icmlaffiliation{sch}{School of Engineering,  University of Glasgow, G12 8QQ, Glasgow, UK}

\icmlcorrespondingauthor{Muhammad Ahmed Mohsin}{muahmed@stanford.edu}
\icmlcorrespondingauthor{Muhammad Umer}{mumer@stanford.edu}
\icmlcorrespondingauthor{Ahsan Bilal}{ahsan.bilal-1@ou.edu}
\icmlcorrespondingauthor{Muhammad Ali Jameshed}{muhammadali.jamshed@glasgow.ac.uk}
\icmlcorrespondingauthor{John M. Cioffi}{cioffi@stanford.edu}

\icmlkeywords{Machine Learning, Wireless Communication, Continual Learning}

\vskip 0.3in
]

\printAffiliationsAndNotice{\icmlEqualContribution}
\begin{abstract}
Modern 5G/6G deployments routinely face \emph{cross-configuration handovers}---users traversing cells with different antenna layouts, carrier frequencies, and scattering statistics---which inflate channel-prediction NMSE by \textbf{37.5\%} on average when models are naively fine-tuned. The proposed improvement frames this mismatch as a continual-learning problem and benchmarks three adaptation families: replay with loss-aware reservoirs, synaptic-importance regularization, and memory-free learning-without-forgetting. Across three representative 3GPP urban micro scenarios, the best replay and regularization schemes cut the high-SNR error floor by up to \textbf{2 dB} ($\approx$35\%), while even the lightweight distillation recovers up to 30\% improvement over baseline handover prediction schemes. These results show that targeted rehearsal and parameter anchoring are essential for handover-robust CSI prediction and suggest a clear migration path for embedding continual-learning hooks into current channel prediction efforts in 3GPP---NR~\cite{polese20183gpp} and O-RAN~\cite{garcia2021ran}. The full codebase can be found at \url{https://github.com/ahmd-mohsin/continual-learning-channel-prediction}.
\end{abstract}

\section{Introduction}
\label{introduction}
Channel state information (CSI) prediction and estimation are long-unsolved problems in wireless communications and a bottleneck to many advanced physical-layer designs. Accurate CSI is essential for multi-antenna systems, but rapid channel variations (``channel aging'') make timely CSI acquisition difficult in practice. 5G network radio (NR) specifications (e.g. TDD mode) mandate uplink sounding only every $\geq$ 2ms~\cite{villena2024aging}. For instance, a 28GHz link with a 60km/h user has a coherence time on the order of 0.3ms~\cite{villena2024aging}, so the channel may drift significantly between pilots. Consequently, outdated CSI can severely degrade throughput: studies show that even a 4ms feedback delay at moderate speed (30km/h) can reduce sum-rate by $\approx$50\% at a carrier frequency of 3.5 GHz~\cite{li2021impact}. In practice, this implies that at higher speeds (e.g. 60km/h) the achievable rate can drop by tens of percent if prediction fails. These facts underscore the critical need for advanced channel prediction to preempt channel aging under 3GPP timing constraints.

Traditional statistical or model-based predictors cannot fully capture real-world channel dynamics, so learning-based methods have been explored~\cite{jiang2019neural}. Recurrent neural networks (RNNs) with Long short-term memory (LSTM)/gated recurrent units (GRU)~\cite{greff2016lstm},~\cite{dey2017gate} and attention-based models (Transformers) can learn to forecast channel time-series from past CSI~\cite{joo2019deep}. However, these data-driven predictors work well only when test channels closely match training conditions. In fact, deep predictors ``exhibit poor generalization, requiring retraining when the CSI distribution changes''~\cite{liu2024llm4cp}. The mismatch in array geometry or mobility can lead to large errors: for example, LSTM/GRU models incur a prediction error of 37.5\% when moving from standard to dense environments (detailed description in Section~\ref{distributions} of the Appendix). Similarly, changing antenna spacing or carrier frequency can worsen the NMSE by on the order of 15-30\% over the nominal case. A simple change of antenna tilt, array spacing, and polarization between different network conditions causes a prediction error of 34\% as Figure~\ref{fig:baseline_dsitributions} shows (see Appendix \ref{app:baseline_results} and Figure \ref{fig:baseline_standard} for the full baseline distributions and zero-shot evaluation). Moreover, sequential multi-step forecasting suffers from error accumulation; small mistakes compound over time, and naïve fine-tuning on new data causes catastrophic forgetting of previous channel patterns. In short, standard deep learning predictors lack cross-configuration generalization, and their sequential operation can amplify errors. This motivates the need for continual adaptation~\cite{kirkpatrick2017overcoming}. The proposed method incorporates Learning without Forgetting (LwF)~\cite{li2017learning} style distillation to preserve older outputs while training on new data.

\begin{figure*}[t!]
\centering
\includegraphics[width=1\textwidth]{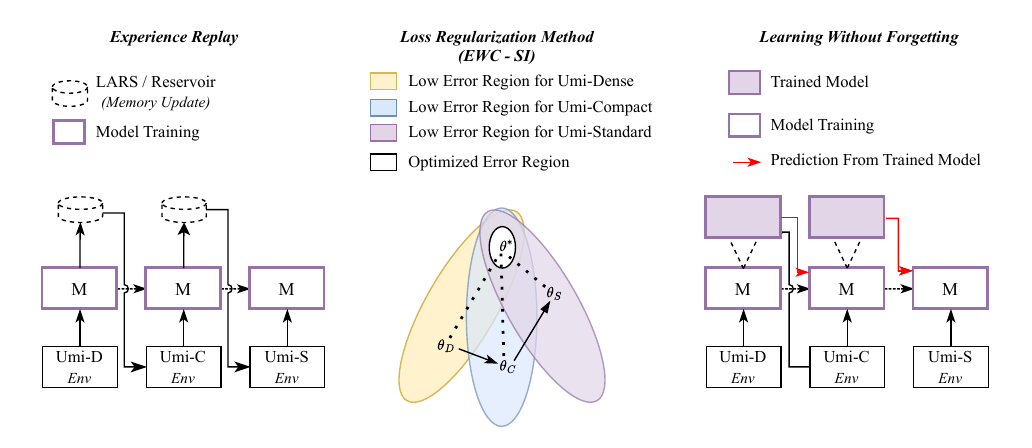}
\caption{Algorithmic flow for continual learning under data drift mismatch for MIMO channel prediction.}
\label{figure:flowchart}
\end{figure*}

\textbf{Contributions.} To address these challenges, we propose a continual learning framework as shown in Figure~\ref{figure:flowchart} that incrementally adapts the channel predictor to evolving network conditions without catastrophic forgetting. This approach integrates replay and regularization techniques while maintaining a buffer of past channel examples for experience replay~\cite{rolnick2019experience}. It also applies methods like Elastic Weight Consolidation (EWC)~\cite{zhou2022elastic} and Synaptic Intelligence~\cite{zenke2017continual} to penalize changes to weights important for old channels. Empirical evaluation shows that Experience Replay (ER), Loss Regularization Method (LRM), and LwF reduce cross-domain channel prediction error by up to 35\%, 32\%, and 25\%, respectively, across diverse network configurations. Furthermore, training a single model continuously over all domains yields an additional 10\% improvement in within-domain prediction accuracy. By incorporating loss-aware experience replay into ER and synaptic intelligence into LRM, we achieve a further average gain of 10\% across datasets. Collectively, this novel continual learning framework delivers robust generalization under cross-network configurations.

\textbf{Related Work.} Data-driven CSI prediction has attracted considerable interest.~\cite{article} proposes an LSTM-based vehicular channel predictor capturing temporal dynamics to outperform ARIMA under a specified velocity profile. ~\cite{jiang2020recurrent} introduced an LSTM predictor for fading channels, exploiting memory for multi-tap channel dependencies, although RNNs suffer from vanishing gradients and limiting lookaheads. \cite{jiang2022accurate} proposed a parallel-attention scheme (JSAC 2022) that forecasts multiple future frames simultaneously. This transformer-based predictor largely eliminates ``mobility-induced'' error but demands extensive matched-condition training and high computational complexity. \cite{zhang2024transformer} extended the attention mechanism to vehicular links: in an RSMA-enabled V2X system they use multi-head attention across subcarriers for CSI prediction, yielding higher rates but at the cost of large model size and training effort. \cite{liu2024llm4cp} explored foundation models to adapt a GPT-2 language model to MISO-OFDM prediction, introducing pre-trained weight sharing for CSI. This enables zero-shot use with minimal fine-tuning but suffers from a natural language and CSI domain gap.

The attention mechanism has been widely used to capture long-range temporal correlation in CSI sequences. \cite{kim2025machine} has explored self-attention to weigh the relevance of different historical pilots when predicting the future channel. Such transformer-based methods can mitigate error propagation over one or two steps, but they still face challenges for long-horizon forecasts. Data-driven channel predictors are known to ``encounter difficulties when dealing with unseen channel conditions''~\cite{kim2025machine}. In concrete terms, a transformer trained on one antenna array will typically see 15-30\% worse NMSE when used on a different array spacing (see Figure \ref{fig:baseline_standard} in Appendix). Domain shift forces SOTA attention models to be retrained or adapted for new configurations, restricting their out-of-the-box applicability.

Generative models---particularly diffusion models---have recently appeared in wireless channel research, mostly for CSI synthesis and augmentation.~\cite{lee2024generatinghighdimensionaluserspecific} condition a diffusion model on UE position to sample new MIMO channel matrices, effectively augmenting a small measured dataset. These synthetic channels can improve tasks like CSI compression or beam selection, as demonstrated in the paper.~\cite{bhattacharya2025sic,zilberstein2024joint} explore Langevin dynamics for joint source channel estimation under MIMO scenarios but face computational time inefficiency. Existing models omit online drift adaptation, motivating our continual learning approach to MIMO channel prediction.

\section{System Model}
\subsection{Dataset} The dataset is synthesized with QuaDRiGa~\cite{jaeckel2014quadriga}, fixing the carrier at 5 GHz, the bandwidth at 100 MHz, and sampling 500 time instants across 18 OFDM resource blocks for each run. The details for the urban micro channel (UMi) are provided in Section~\ref{distributions} of the Appendix (see Table \ref{tab:umi-uma-scenarios} for parameter settings and Figure \ref{fig:net_config_umi_uma} for channel‐gain distributions). For every Monte-Carlo seed, 256 users have randomly chosen azimuths and configuration-dependent ranges; then they initialize linear tracks that match QuaDRiGa's sample-density constraints. To overcome the simulator's seeding artifacts and expose temporal correlation, users are displaced slightly at each iteration so successive channels remain correlated sand the predictor must infer the evolution to the next state.

\subsection{Model Configuration} \label{models}
For continual-learning channel prediction, we test 3 different machine learning models to evaluate the robustness of our proposed pipeline and achieve accurate results. LSTM~\cite{greff2016lstm} performs the best overall across all datasets. Gated recurrent unit (GRU)~\cite{dey2017gate} and Transformer~\cite{han2021transformer} both perform equally well under different scenarios. Due to space constraints, complete architectural diagrams and hyperparameter settings for the LSTM, GRU and Transformer backbones appear in Appendix \ref{app:baseline_results}.

\section{Continual Learning for Channel Prediction}
Dynamic wireless channels induce distribution shift (\emph{data drift}; see Section~\ref{introduction}), so a predictor trained once quickly becomes stale across network configurations. Continual learning offers a remedy: given a sequence of datasets $\mathcal{D}_1,\mathcal{D}_2,\dots$, the model updates on $\mathcal{D}_{k}$ while regularizing against weight changes that would degrade performance on $\bigcup_{i<k}\mathcal{D}_{i}$. Without such constraints, na\"ive sequential training biases the weights toward the most recent distribution, causing catastrophic forgetting on earlier conditions and yielding sub-optimal performance overall.

\subsection{Experience Replay} \label{sec:er}
ER is a powerful tool in continual learning where on-policy learning from novel instances (current dataset) and off-policy learning~\cite{maei2010toward} from replay experiences (previous dataset) are interleaved. ER replays past channel samples to mitigate forgetting under dynamic wireless configurations. This reduces catastrophic forgetting~\cite{kemker2018measuring} and allows the same channel prediction models to be continuously used on real-time datasets gathered from different environments for better efficacy.

A replay buffer, $\mathcal{M}$, with a defined maximum capacity $N_{\text{buffer}}$, serves as the repository for these experiences. When a new experience $e_t$ occurs at time $t$ (corresponding to a new data sample from a specific network configuration), a decision is made regarding its inclusion in $\mathcal{M}$ based on a predetermined strategy. A user equipment (UE) that roams across several cells experiences \emph{task shifts} in the underlying propagation environment---e.g.\ UMi-\textit{compact} $\rightarrow$ UMi-\textit{dense}
$\rightarrow$ UMi-\textit{standard}, where UMi stands for Urban microcell. To maintain reliable link adaptation, the channel-state predictor must acquire the current cell's statistics \emph{without catastrophically forgetting} the statistics learned in previous cells' network configurations. We follow the continual-learning paradigm of \emph{experience replay} to achieve this trade-off~\cite{liu2023wireless}.

\textbf{Replay buffer.} Algorithm~\ref{alg:replay} allocates a fixed-size memory $\mathcal{M}=\!\{\!e^{(n)}\!\}_{n=1}^{N_{\text{buffer}}}$
that stores past observations,
$
  e^{(n)} \;=\;
\bigl(
  \mathbf{X}^{(n)},          % input sequence
  \mathbf{H}^{(n)},          % ground-truth next-step channel
  \phi^{(n)}                 % optional meta-data (SNR, scenario id, …)
\bigr)
$
where $\mathbf{X}^{(n)}\!\in\!\mathbb{C}^{2\times T\times N_\mathrm{tx}\times N_\mathrm{rx}}$
is a sequence of $T$ past channel realizations (real/imag split) and $\mathbf{H}^{(n)}$ is the target channel matrix to be predicted. Throughout training we denote by $\mathcal{D}_{k}$ the mini-dataset collected in the \mbox{$k$-th} cell.

\textbf{Mini-batch composition.} Global training step~$t$ (UE currently in cell~$k$) draws a mini-batch~\cite{krutsylo2024batch}
$\mathcal{B}_{t}
   = \mathcal{B}_{\text{current}}\cup\mathcal{B}_{\text{replay}}$
where $\mathcal{B}_{\text{current}}\!\subset\!\mathcal{D}_{k}$ and $\mathcal{B}_{\text{replay}}\!\subset\!\mathcal{M}$. Elements of $\mathcal{B}_{\text{replay}}$ act as \emph{rehearsal anchors} that remind the network of previous propagation settings. Mathematically, the training process with experience replay can be formulated as follows. Let $\bm{\theta}$ represent the parameters of the channel prediction model. The loss function associated with the current task, $\mathcal{L}_{\text{current}}(\bm{\theta}, \mathcal{B}_{\text{current}})$, is computed based on the model's predictions on the data from the current network configuration. When experience replay is employed, a mini-batch of past experiences, $\mathcal{B}_{\text{replay}}$, is sampled from the replay buffer $\mathcal{M}$. A loss term, $\mathcal{L}_{\text{replay}}(\bm{\theta}, \mathcal{B}_{\text{replay}})$, is then calculated based on the model's performance on this replayed data. The overall loss function for a given training step can be a weighted combination of these two loss terms~\cite{fujimoto2020equivalence}:
\begin{equation}
\mathcal{L}_{\text{total}}(\bm{\theta}) = \lambda \mathcal{L}_{\text{current}}(\bm{\theta}, \mathcal{B}_{\text{current}}) + (1-\lambda) \mathcal{L}_{\text{replay}}(\bm{\theta}, \mathcal{B}_{\text{replay}})
\end{equation}
where $\lambda \in [0,1]$ is a hyperparameter that governs the balance between learning from the current task and rehearsing past experiences. Channel prediction specifically employs the normalized mean square error (NMSE) as the loss function. For the current dataset, this is:
\begin{figure}[t!]
    \centering
    \includegraphics[width=\linewidth]{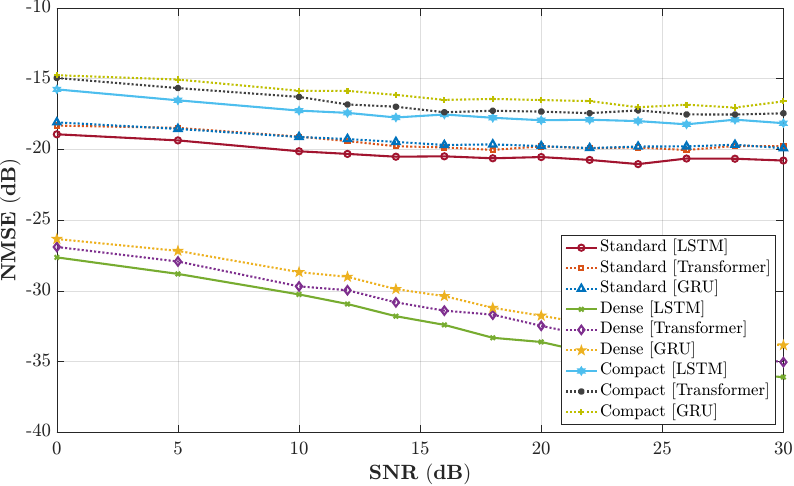}
    \caption{Baseline SNR trained on UMi dense and tested on all scenarios under all architectures.}
    \label{fig:baseline_dsitributions}
\end{figure}
\begin{equation}
\mathcal{L}_{\text{current}}(\bm{\theta}, \mathcal{B}_{\text{current}}) = \frac{1}{|\mathcal{B}_{\text{current}}|} \sum_{i=1}^{|\mathcal{B}_{\text{current}}|} \frac{\|\mathbf{H}_i - \hat{\mathbf{H}}_i(\bm{\theta})\|_F^2}{\|\mathbf{H}_i\|_F^2}
\end{equation}
where $\mathbf{H}_i$ represents the true channel matrix for the $i$-th sample in $\mathcal{B}_{\text{current}}$, $\hat{\mathbf{H}}_i(\bm{\theta})$ is the predicted channel matrix using parameters $\bm{\theta}$, and $\|\cdot\|_F$ denotes the Frobenius norm. Similarly, for the replayed experiences, the loss is:
\begin{equation}
\mathcal{L}_{\text{replay}}(\bm{\theta}, \mathcal{B}_{\text{replay}}) = \frac{1}{|\mathcal{B}_{\text{replay}}|} \sum_{j=1}^{|\mathcal{B}_{\text{replay}}|} \frac{\|\mathbf{H}_j - \hat{\mathbf{H}}_j(\bm{\theta})\|_F^2}{\|\mathbf{H}_j\|_F^2}
\end{equation}

The complete training objective with experience replay thus becomes:
\begin{align}
\label{eq:total_loss}
\mathcal{L}_{\text{total}}(\bm{\theta}) &= \lambda \left( \frac{1}{|\mathcal{B}_{\text{current}}|} \sum_{i=1}^{|\mathcal{B}_{\text{current}}|} \frac{\|\mathbf{H}_i - \hat{\mathbf{H}}_i(\bm{\theta})\|_F^2}{\|\mathbf{H}_i\|_F^2} \right) \\
&+ (1-\lambda) \left( \frac{1}{|\mathcal{B}_{\text{replay}}|} \sum_{j=1}^{|\mathcal{B}_{\text{replay}}|} \frac{\|\mathbf{H}_j - \hat{\mathbf{H}}_j(\bm{\theta})\|_F^2}{\|\mathbf{H}_j\|_F^2} \right)
\end{align}

The mixing ratio $\lambda$ is another crucial hyperparameter~\cite{fedus2020revisiting}. Larger $\lambda$ biases optimization toward the current configuration’s loss, driving parameters away from prior-environment optima and thus causing catastrophic forgetting. Conversely, a lower value of $\lambda$ places more emphasis on the replayed data, encouraging the model to retain knowledge from past tasks but potentially hindering its ability to learn new patterns from the current data. In this 3-task study ($\text{UMi}:\{\text{compact},\text{dense},\text{standard}\}$) we fix $N_{\text{buf}}=5000\;\text{samples}\approx 10$MB, small enough to be cached on the gNB side. Furthermore, high $\lambda$ expedites adaptation after handover but risks NMSE spikes when the UE returns to a previous cell; low $\lambda$ yields smoother performance across cells at the cost of slower convergence.

\begin{algorithm}[tb]
\caption{Continual Channel Prediction with Experience Replay (Reservoir -- LARS)}
\label{alg:replay}
\begin{algorithmic}[1]
\STATE {\bfseries Input:} buffer size $N_{\mathrm{buf}}$, mixing weight $\lambda$, sampling mode $s\in\{\textsc{Uniform},\textsc{LARS}\}$, learning rate $\eta$
\STATE Initialize replay buffer $\mathcal{M}\leftarrow\emptyset$, counter $t\leftarrow 0$, model parameters $\theta$
\FUNCTION{\texttt{Insert}($\mathbf{X}$, $\mathbf{H}$, $\ell$)}
  \STATE $t \leftarrow t + 1$
  \IF{$|\mathcal{M}| < N_{\mathrm{buf}}$}
    \STATE $\mathcal{M} \leftarrow \mathcal{M} \cup \{(\mathbf{X},\mathbf{H},\ell)\}$
  \ELSIF{$\text{rand}() < \frac{N_{\mathrm{buf}}}{t}$}
    \IF{$s = \textsc{LARS}$}
      \STATE choose victim $v$ using Eq.~(7)
    \ELSE
      \STATE $v \leftarrow \text{randint}(1, |\mathcal{M}|)$
    \ENDIF
    \STATE $\mathcal{M}[v] \leftarrow (\mathbf{X},\mathbf{H},\ell)$
  \ENDIF
\ENDFUNCTION
\FOR{each visited cell $k$ \textbf{do}}
  \FOR{each measurement $(\mathbf{X}_t,\mathbf{H}_t)$ \textbf{do}}
    \STATE $\hat{\mathbf{H}}_t \leftarrow f_\theta(\mathbf{X}_t)$
    \STATE $\ell_t \leftarrow \ell_{\text{NMSE}}(\mathbf{H}_t,\hat{\mathbf{H}}_t)$
    \STATE \texttt{Insert}($\mathbf{X}_t$, $\mathbf{H}_t$, $\ell_t$)
    \IF{ready to update}
      \STATE sample $\mathcal{B}_{\text{curr}} \subset \mathcal{D}_k$, $\mathcal{B}_{\text{rep}} \subset \mathcal{M}$
      \STATE compute $\mathcal{L}_{\text{total}}$ via Eq.~(4)
      \STATE $\theta \leftarrow \theta - \eta\nabla_\theta\mathcal{L}_{\text{total}}$
    \ENDIF
  \ENDFOR
\ENDFOR
\end{algorithmic}
\end{algorithm}

% \begin{algorithm}[t]
% \caption{ER (Reservoir-LARS) for Continual CSI Prediction}
% \label{alg:replay}
% \begin{algorithmic}[1]
% \Require $N_{\mathrm{buf}},\,\lambda,\,s\in\{\mathrm{U},\mathrm{L}\},\,\eta$
% \State $\mathcal M\!\leftarrow\!\emptyset,\;t\!\leftarrow\!0,\;\theta\!\leftarrow\!\theta_0$
% \For{$k\!\in\!\{\text{cells}\}$}
%   \For{$(X_t,H_t)$}
%     \State $\hat H_t = f_\theta(X_t),\;\ell_t = \ell_{\mathrm{NMSE}}(H_t,\hat H_t),\;t\leftarrow t+1$
%     \If{$|\mathcal M|<N_{\mathrm{buf}}$}
%       \State $\mathcal M\cup\{(X_t,H_t,\ell_t)\}$
%     \ElsIf{$u\sim U(0,1)<\frac{N_{\mathrm{buf}}}{t}$}
%       \State $v\sim\begin{cases}
%         \mathrm{LARS}(\{\ell_i\}_{i=1}^{|\mathcal M|}), & s=\mathrm{L}\\
%         \mathrm{Unif}\{1,\dots,|\mathcal M|\}, & s=\mathrm{U}
%       \end{cases}$
%       \State $\mathcal M_v \leftarrow (X_t,H_t,\ell_t)$
%     \EndIf
%     \If{update}
%       \State $\mathcal B_c\subset D_k,\;\mathcal B_r\subset \mathcal M$
%       \State $\mathcal L_{\mathrm{tot}}=\lambda\,\mathcal L_{\mathrm{curr}}+(1-\lambda)\,\mathcal L_{\mathrm{rep}}$
%       \State $\theta\leftarrow\theta-\eta\,\nabla_\theta\mathcal L_{\mathrm{tot}}$
%     \EndIf
%   \EndFor
% \EndFor
% \end{algorithmic}
% \end{algorithm}

\textbf{Reservoir Sampling.} In the continual-learning pipeline, the UE encounters a \textit{stream} of channel measurements while traversing the \mbox{UMi-\{compact, dense, standard\}} layouts. At step $t$ Algorithm~\ref{alg:replay} observes the pair $e_t = (\mathbf{X}_t,\mathbf{H}_t)$, where $\mathbf{X}_t\!\in\!\mathbb{C}^{2\times T\times N_\mathrm{tx}\times N_\mathrm{rx}}$ is the window of the \emph{past} $T$ channel snapshots (real/imaginary split) and $\mathbf{H}_t\!\in\!\mathbb{C}^{2\times N_\mathrm{tx}\times N_\mathrm{rx}}$ is the \emph{next-slot} channel to be predicted. We maintain a replay buffer $\mathcal{M}$ of fixed cardinality $N_{\text{buffer}}$, physically cachable at the serving gNB (5G base station)~\cite{kim2020imbalanced}. To guarantee that every measurement---whether collected in the first or the last visited cell---has the same chance of being rehearsed, Algorithm~\ref{alg:replay} adopts classical \emph{reservoir sampling}:
\begin{equation}
  \Pr\!\bigl[e_i\!\in\!\mathcal{M}_t\bigr]
  =\frac{N_{\text{buffer}}}{t},
  \quad
  i=1,\dots,t,
\end{equation}
where $\mathcal{M}_t$ denotes the buffer content after $t$ total observations. The procedure is cell-agnostic and thus well-suited to heterogeneous mobility traces.
\begin{enumerate}[leftmargin=*]
  \item \textbf{Data acquisition.}
        While the UE resides in scenario~$k$
        (e.g.\ UMi-dense), incoming measurement
        $e_t$ is fed to the reservoir algorithm,
        which either stores it or discards it
        with probability $1-\tfrac{N_{\text{buffer }}}{t}$.
  \item \textbf{Mini-batch assembly.}
        Each stochastic gradient descent (SGD) step draws a batch  
        $\mathcal{B}_{\text{curr}}\!\subset\!\mathcal{D}_{k}$
        from the \emph{live} cell trace
        and a rehearsal batch  
        $\mathcal{B}_{\text{rep}}\!\subset\!\mathcal{M}$,
        then optimizes the mixed NMSE loss
        of Eq.~\eqref{eq:total_loss}.
  \item \textbf{UE hand-over.}
        When the UE moves to the next cell
        ($k\!\!\to\!k\!+\!1$),
        the same buffer $\mathcal{M}$ is retained, guaranteeing that
        legacy UMi statistics remain rehearsed even if those
        environments are no longer observed.
\end{enumerate}
\begin{algorithm}[tb]
\caption{Continual Channel Prediction with EWC}
\label{alg:ewc}
\begin{algorithmic}[1]
\STATE {\bfseries Input:} learning rate $\eta$, stability coefficient $\alpha$
\STATE Initialize model parameters $\theta$ 
\STATE Initialize bank of snapshots $\mathcal{B}\leftarrow\emptyset$ \COMMENT{stores $(\theta_j^{*},F_j)$}
\FOR{each task $D_k$ (\textit{UMi-compact} $\rightarrow$ \textit{dense} $\rightarrow$ \textit{standard}) \textbf{do}}
  \FOR{each mini-batch $(\mathbf{X},\mathbf{H}) \in D_k$ \textbf{do}}
    \STATE $\hat{\mathbf{H}} \leftarrow f_\theta(\mathbf{X})$
    \STATE $\ell_{\text{NMSE}} \leftarrow \ell_{\text{NMSE}}(\mathbf{H},\hat{\mathbf{H}})$
    \STATE $\ell_{\text{EWC}} \leftarrow \frac{\alpha}{2} \sum_{(\theta_j^{*},F_j)\in\mathcal{B}} \sum_i F_{j,i}(\theta_i-\theta_{j,i}^{*})^{2}$
    \STATE $\theta \leftarrow \theta - \eta\nabla_\theta(\ell_{\text{NMSE}} + \ell_{\text{EWC}})$
  \ENDFOR
  \STATE Fisher computation for $D_k$:
  \STATE $F_{k,i} \leftarrow \frac{1}{|D_k|} \sum_{(\mathbf{X},\mathbf{H})\in D_k} (\partial_{\theta_i}\ell_{\text{NMSE}})^{2}$
  \STATE Store snapshot: $\theta_k^{*} \leftarrow \theta$, $\mathcal{B} \leftarrow \mathcal{B} \cup \{(\theta_k^{*},F_k)\}$
\ENDFOR
\end{algorithmic}
\end{algorithm}
\textbf{Loss-Aware Reservoir Sampling (LARS).} While uniform reservoir sampling treats all past observations equally, cell-edge channels or deep-fade events (precisely the cases that most hurt throughput) may be under-represented. Algorithm~\ref{alg:replay} therefore adopts \emph{loss-aware reservoir sampling} (LARS)~\cite{mall2023change, kumari2022retrospective}, which biases the buffer in favor of those channel realizations with which the predictor still struggles. Immediately after performing the forward pass on mini-batch $\mathcal{B}_{t}$, Algorithm~\ref{alg:replay} computes the per-observation NMSE $\mathcal{L}_{\text{current}}(\bm{\theta}, \mathcal{B}_{\text{current}})$
and stores it alongside $e_i$ if the item enters the buffer. When the buffer $\mathcal{M}$ is full ($|\mathcal{M}| = N_{\text{buffer}}$) a newly arrived observation $e_t$ is considered for inclusion with the same reservoir probability $\tfrac{N_{\text{buffer}}}{t}$. If the decision is \textit{keep}, LARS chooses a \emph{victim} index $v$ according to
\begin{equation}
  \Pr[v=i] \;=\;
  \frac{\bigl(\ell_i + \epsilon\bigr)^{-1}}
       {\displaystyle
        \sum_{j=1}^{N_{\text{buf}}} (\ell_j + \epsilon)^{-1}},
  \label{eq:lars_prob}
\end{equation}
where \(\epsilon\!>\!0\) prevents division by zero. Hence observations whose NMSE has already dropped are \emph{more} likely to be evicted, whereas hard-to-predict channels (e.g.\ severe multi-path or rich scattering) persist longer in $\mathcal{M}$. Retaining difficult samples ensures that the predictor keeps rehearsing rare but performance-critical propagation states (deep fades, high delay spreads, cell-edge SNRs).

\begin{figure*}[t!]
  \centering
  % ---------- first sub-figure ----------
  \begin{subfigure}[t]{0.32\textwidth}
      \centering
      \includegraphics[width=\textwidth]{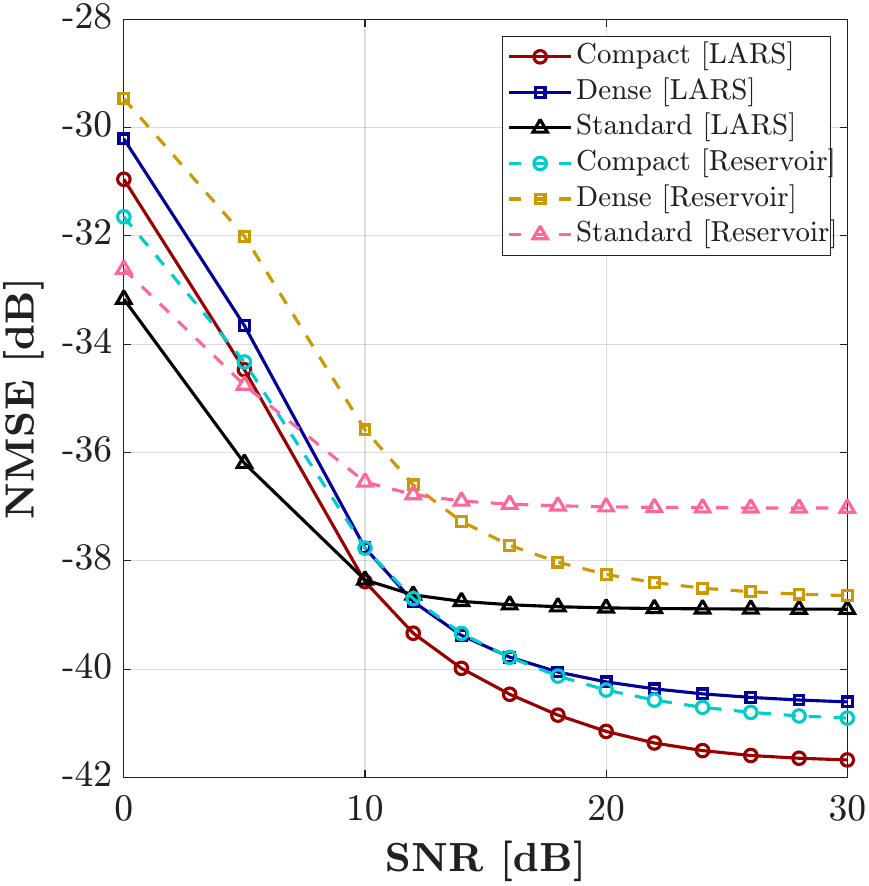}
      \caption{SNR [dB] vs. NMSE [dB] for experience replay with reservoirs and LARS.}
      \label{fig:experience_replay}
  \end{subfigure}%
  \hspace{0.5em}
  % ---------- second sub-figure ----------
  \begin{subfigure}[t]{0.32\textwidth}
      \centering
      \includegraphics[width=\textwidth]{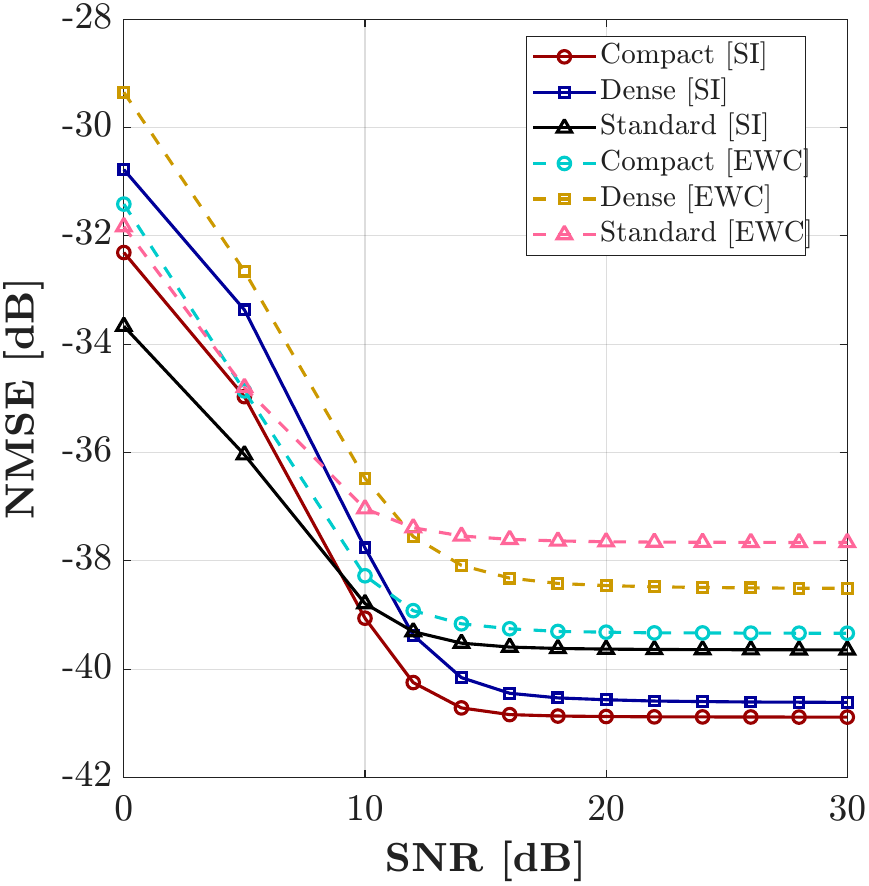}
      \caption{SNR [dB] vs. NMSE [dB] for loss regularization method for SI and EWC.}
      \label{fig:loss_regularization}
  \end{subfigure}%
  \hspace{0.5em}
  \begin{subfigure}[t]{0.32\textwidth}
      \centering
      \includegraphics[width=\textwidth]{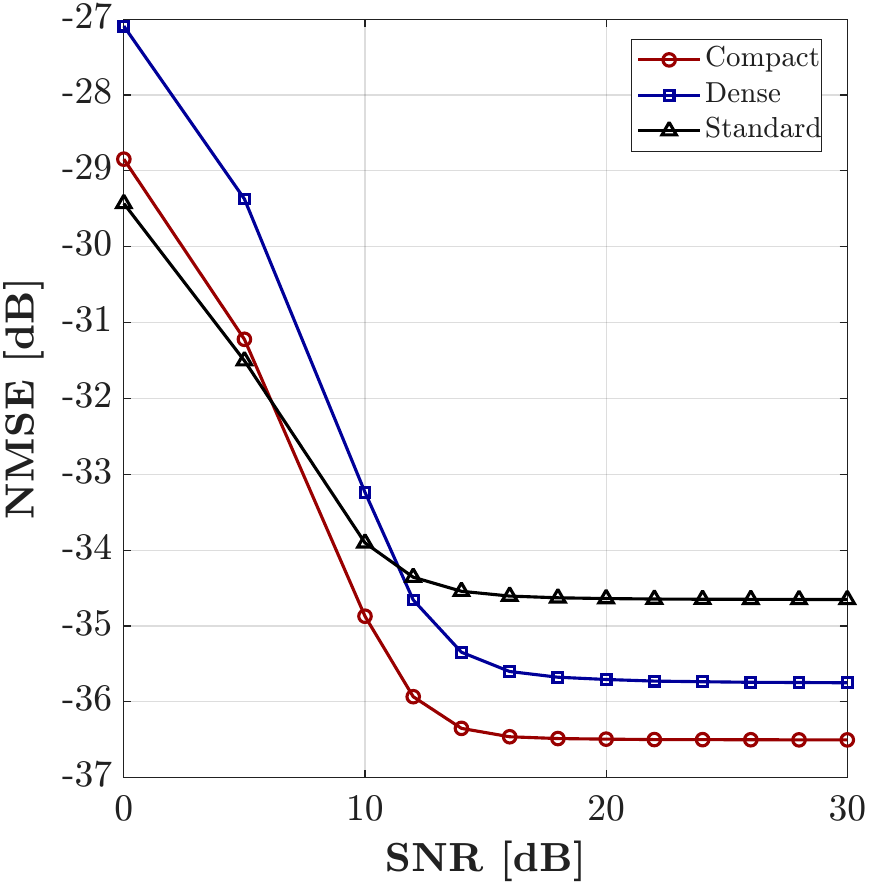}
      \caption{SNR [dB] vs. NMSE [dB] for LWF.}
      \label{fig:ewf}
  \end{subfigure}%
  \hspace{0.5em}
  \caption{SNR vs NMSE curves for various continual learning methods tested for cross-network generalization.}
  \label{fig:continual_learning}
\end{figure*}

\subsection{Loss Regularization Method}
Augmenting the training loss with penalty terms prevents catastrophic forgetting by discouraging large updates to parameters critical for prior tasks~\cite{zhao2024statistical}. We adopt two complementary regularizers: EWC~\cite{yang2021elastic}, which imposes a Fisher-information-weighted quadratic penalty~\cite{calmet2005dynamics}, and SI~\cite{zenke2017continual}, which dynamically accumulates per-weight importance from loss gradients to penalize significant updates, preserving performance across the UMi sequence (\{\text{compact}, \text{dense}, \text{standard}\}).

\begin{algorithm}[tb]
\caption{Continual Channel Prediction with Learning-without-Forgetting}
\label{alg:lwf}
\begin{algorithmic}[1]
\REQUIRE mixing weight $\lambda$, learning rate $\eta$, datasets $\{\mathcal{D}_k\}_{k=1}^K$
\STATE Initialize model parameters $\theta$
\STATE Train on $\mathcal{D}_1$ by minimizing:
\STATE $\mathcal{L}_{\text{task}}(\theta) = \frac{1}{|\mathcal{D}_1|}\sum_{(\mathbf{X},\mathbf{H})\in\mathcal{D}_1} \frac{\|\mathbf{H} - f_\theta(\mathbf{X})\|_F^2}{\|\mathbf{H}\|_F^2}$
\STATE $\theta_{\text{old}} \leftarrow \theta$
\FOR{$k = 2$ \textbf{to} $K$}
  \FOR{\textbf{all} minibatch $(\mathbf{X},\mathbf{H}) \subset \mathcal{D}_k$}
    \STATE $\hat{\mathbf{H}} \leftarrow f_\theta(\mathbf{X})$
    \STATE $\hat{\mathbf{H}}_{\text{old}} \leftarrow f_{\theta_{\text{old}}}(\mathbf{X})$
    \STATE $\mathcal{L}_{\text{task}} \leftarrow \frac{1}{|\mathcal{B}|}\sum \frac{\|\mathbf{H} - \hat{\mathbf{H}}\|_F^2}{\|\mathbf{H}\|_F^2}$
    \STATE $\mathcal{L}_{\text{KD}} \leftarrow \frac{1}{|\mathcal{B}|}\sum \frac{\|\hat{\mathbf{H}}_{\text{old}} - \hat{\mathbf{H}}\|_F^2}{\|\hat{\mathbf{H}}_{\text{old}}\|_F^2}$
    \STATE $\mathcal{L} \leftarrow \mathcal{L}_{\text{task}} + \lambda\,\mathcal{L}_{\text{KD}}$
    \STATE $\theta \leftarrow \theta - \eta\,\nabla_\theta\,\mathcal{L}$
  \ENDFOR
  \STATE $\theta_{\text{old}} \leftarrow \theta$
\ENDFOR
\end{algorithmic}
\end{algorithm}
\textbf{Elastic Weight Consolidation (EWC).} \label{sec:ewc}
 When the UE goes into a given UMi cell, Algorithm~\ref{alg:ewc} treats the resulting mini-dataset $D_k=\{(\mathbf{X}^{(n)},\mathbf{H}^{(n)})\}_{n=1}^{N_k}$ as one \emph{task}. After training on $D_k$, Algorithm~\ref{alg:ewc} obtains the weight vector $\bm{\theta}_k^{*}$ that minimises the NMSE on that cell's pathloss and fading statistics.

To achieve this, EWC identifies which weights must remain stable by measuring how sensitive the NMSE is to each parameter. Specifically, it estimates the importance of each parameter $\theta_i$ for the $k^{\text{th}}$ task using an approximation of the diagonal of the Fisher information matrix $F(k, i)$: 
\begin{equation} \label{eq:fisher_eq}
  F_{k,i}
  \;=\;
  \frac{1}{|D_k|}
  \sum_{(\mathbf{X},\mathbf{H})\in D_k}
  \Bigl(
     \partial_{\theta_i}\,
     \mathcal{L}_{\mathrm{NMSE}}\bigl(\bm{\theta}_k^{*};\mathbf{X},\mathbf{H}\bigr)
  \Bigr)^{\!2}
\end{equation}

The expression inside parenthesis, \(\partial_{\theta_i}\,\mathcal{L}_{\mathrm{NMSE}}(\bm{\theta}_k^{*};\mathbf{X},\mathbf{H})\), is the gradient of the NMSE loss with respect to the parameter \(\theta_i\), reflecting how sensitively a small change in \(\theta_i\) perturbs the characteristics of cell \(k\).

When the UE hands over to the next propagation scenario $D_{k+1}$, Algorithm~\ref{alg:ewc} introduces a quadratic penalty that keeps $\bm{\theta}$ near $\bm{\theta}^k$ proportionally to $F{(k,i)}$. For a single previous task, the EWC regularization term is
\begin{equation}
    \mathcal{L}_{\text{EWC}}(\bm{\theta}) = \frac{\alpha}{2} \sum_{i} F_{k,i} \left( \theta_i - \theta^{k,i} \right)^2
\end{equation}
where $\alpha>0$ is a stability coefficient ($\alpha=0.4$) that balances how strongly EWC penalizes deviation from task $k$'s optimum. After several cell visits the UE has encountered $\mathcal{T}_{1{:}k}=\{$UMi-\textit{compact}, UMi-\textit{dense}, UMi-\textit{standard}$\}$. Algorithm~\ref{alg:ewc} maintains a \emph{bank} of snapshots $\{(\bm{\theta}_j^{*},F_j)\}_{j=1}^{k}$.

The penalty therefore generalizes to:
\begin{equation}\label{eq:ewc_multi}
  \mathcal{L}_{\text{EWC}}(\bm{\theta})
  \;=\;
  \frac{\alpha}{2}\sum_{i}\!
  \sum_{j=1}^{k}
  F_{j,i}\bigl(\theta_i-\theta_{j,i}^{*}\bigr)^{2}.
\end{equation}
Equation \eqref{eq:ewc_multi} encourages the predictor to \emph{retain the parameters that capture the key statistical properties of the channel distribution in each environment}, such as those characterizing the compact (canyon-like), dense (urban block), and standard (wide street) scenarios, while still allowing less-critical weights to adapt to the new cell's Doppler or fading patterns.

In conclusion, after training on task \(D_k\), Algorithm~\ref{alg:ewc} computes Fisher information \(F_k\) from NMSE gradients and stores \(\bm{\theta}^*_k\) alongside \(F_k\). Algorithm~\ref{alg:ewc} applies aggregated Fisher penalties to preserve parameters vital for past channel conditions while adapting to new fading dynamics. Since \(F_{j,i}\) derives directly from NMSE gradients, it naturally targets weights governing high-energy taps, LOS components, and dominant eigenmode features essential for accurate CQI.

\textbf{Synaptic Intelligence (SI).} SI dynamically tracks parameter importance during training without requiring an explicit Fisher computation. Fisher-based weighting often fails in practice, 1) doubles training time with a full extra dataset pass, 2) exhausts GPU memory storing per-parameter importances, and 3) relies on a quadratic local-curvature approximation that becomes unreliable when mini-batch gradients exhibit high stochasticity~\cite{van2025computation},~\cite{puiu2022rethinking}. To achieve a more robust, continual learning strategy, results therefore compare this Fisher approach against SI. SI avoids all three drawbacks by accumulating importance continuously from the same gradients already used for optimization. These drawbacks are magnified in the UMi sequence, so a second pass would double training time; the diagonal Fisher for $\sim\!15$ M parameters would exceed the 24 GB memory budget of a single A6000 GPU. For the $\sim\!400$ UMi channel states, EWC is untenable: each diagonal Fisher for $15$ M parameters ($15\times10^{6}\!\times\!4$ B $\approx$ 60 MB) costs an extra epoch, and storing one per state ($400\times60$ MB $\approx$ 24 GB) already fills an A6000’s 24 GB.

By accumulating parameter importance online from the same gradients used for optimization, incurring only one 32\,bit float ($\sim4$\,bytes) per weight and requiring no extra data sweeps, SI lowers the high-SNR NMSE floor by approximately 0.8--1.4\,dB ($\approx10\%$ MSE reduction) relative to EWC, making it readily scalable through the UMi-compact $\rightarrow$ UMi-dense $\rightarrow$ UMi-standard pipeline~\cite{zenke2017continual}. SI assigns to each parameter $\theta_i$ a ``synaptic importance'' $\omega_i$ based on how much change in $\theta_i$ contributes to reducing the loss on a task, as shown in Algorithm~\ref{alg:si}. During minibatch $D_k$ (e.g.\ the UE dwelling in a specific UMi cell), we track, for every weight $\theta_i$, how much that weight actually \emph{helps} reduces the NMSE  on the instantaneous channel pair $(\mathbf{X},\mathbf{H})$:
\begin{equation}
  \tilde{\omega}_i {=}\; \tilde{\omega}_i \;{+} 
  \bigl(\nabla_{\!\theta_i}\mathcal{L}_{\text{NMSE}}\bigr)^{2}\,\eta
\end{equation}
where $\nabla_{\theta_i}\mathcal{L}_{NMSE}$ is the gradient of the instantaneous mean-squared error loss with respect to $\theta_i$, which normalizes the Frobenius error by $\|\mathbf{H}\|_F^2$, reflecting the link-adaptation metric used by the gNB scheduler. This accumulation $\tilde{\omega}_i$ intuitively measures the total contribution of $\theta_i$ to loss reduction during task $D_k$.

After completing training on $D_k$, let $\Delta \theta_i = \theta_i - \theta_i^{(0)}$ be the total change in parameter $i$ during this task. Algorithm~\ref{alg:si} then finalizes the importance of $\theta_i$ for task $k$ by updating 
\begin{equation} 
  \omega_i \;{+}{=}\;
  \frac{\tilde{\omega}_i}{(\Delta\theta_i)^{2}+\xi},
  \quad
  \Delta\theta_i=\theta_i-\theta_i^{(0)}.
\end{equation} 

The above update increases $\omega_i$ significantly if a large accumulated gradient (large $\tilde{\omega}_i$) managed to cause only a small net change $(\Delta \theta_i)$, indicating that $\theta_i$ was repeatedly pulled by the loss (i.e., in the case of deep fade samples) but resisted changing, a sign that $\theta_i$ is important for maintaining performance, giving evidence that $i$ governs the core fading statistics. We also set $\theta_i^{(0)} \leftarrow \theta_i$ (forward final weights as the reference for the next task $D_{k+1}$). All $\omega_i$ values persist across cells, so the model remembers which parameters matter for, say, the rich scattering UMi-dense layout even after roaming into a compact canyon.

Given the reference parameters $\theta^{(0)}$ from the start of the current task (which equal the parameters after $D_k$), the SI loss for task $D_{k+1}$ is formulated as: 
\begin{equation} 
  \mathcal{L}_{\text{SI}}(\bm{\theta})
  \;=\;
  \frac{\beta}{2}
  \sum_{i}\omega_i\,
  \bigl(\theta_i-\theta_i^{(0)}\bigr)^{2},
  \quad
  \beta=0.6.
\end{equation}
\begin{algorithm}[tb]
\caption{Continual Channel Prediction with Synaptic Intelligence (SI)}
\label{alg:si}
\begin{algorithmic}[1]
\REQUIRE learning rate $\eta$, SI weight $\beta$, damping $\xi$
\STATE Initialize for all $i$:
\STATE $\theta_i$, $\omega_i \leftarrow 0$, $\tilde{\omega}_i \leftarrow 0$, $\theta_i^{(0)} \leftarrow \theta_i$
\FOR{each task $D_k$ (UMi \textit{compact} $\rightarrow$ \textit{dense} $\rightarrow$ \textit{standard}) \textbf{do}}
  \FOR{each mini-batch $(X,H) \in D_k$ \textbf{do}}
    \STATE $\hat{H} \leftarrow f_\theta(X)$
    \STATE $\ell \leftarrow \ell_{\text{NMSE}}(H,\hat{H})$
    \STATE $g_i \leftarrow \nabla_{\theta_i}\ell$
    \STATE $\theta_i \leftarrow \theta_i - \eta\,g_i \quad \forall i$
    \STATE $\tilde{\omega}_i \;{+}{=}\; g_i^2\,\eta \quad \forall i$
  \ENDFOR
  \FOR{each parameter $i$ \textbf{do}}
    \STATE $\Delta\theta_i \leftarrow \theta_i - \theta_i^{(0)}$
    \STATE $\omega_i \;{+}{=}\; \frac{\tilde{\omega}_i}{(\Delta\theta_i)^2 + \xi}$ \COMMENT{SI update}
    \STATE $\tilde{\omega}_i \leftarrow 0, \theta_i^{(0)} \leftarrow \theta_i$
  \ENDFOR
\ENDFOR
\STATE Add SI penalty to the loss:
\STATE $\mathcal{L} = \ell + \frac{\beta}{2}\sum_{i}\omega_i(\theta_i - \theta_i^{(0)})^{2}$
\end{algorithmic}
\end{algorithm}
with $\beta>0$ a weighting hyperparameter (Algorithm~\ref{alg:si} uses $\beta=0.6$) analogous to $\alpha$ in EWC. In wireless channel adaptation, this penalty scales each update to $\theta_i$ by its importance~$\omega_i$, preventing drift from values learned on prior UMi channel conditions. SI accumulates the product of each parameter update and its instantaneous gradient to compute $\omega_i$ online with an $\mathcal{O}(1)$ memory footprint per weight, thereby preferentially regularizing parameters whose sustained gradient magnitudes encode dominant path-loss attenuation and delay-spread dynamics.
\begin{table*}[h!]
  \centering
  \caption{\textbf{Evaluation loss comparison under dynamic continual learning pipelines (sequence length = 32, replay memory size = 5000) [NMSE Loss in dB]}}
  \label{tab:evaluation_losses}
  \vspace{2pt}
  \scalebox{0.8}{
  \begin{tabular}{l|ccc|ccc|ccc}
    \hline
    \textbf{Continuous Learning Pipelines}
      & \multicolumn{3}{c|}{\textbf{Test: UMi Compact}}
      & \multicolumn{3}{c|}{\textbf{Test: UMi Dense}}
      & \multicolumn{3}{c}{\textbf{Test: UMi Standard}} \\
    \cline{2-10}
      & \textbf{Trans.} & \textbf{LSTM} & \textbf{GRU}
      & \textbf{Trans.} & \textbf{LSTM} & \textbf{GRU}
      & \textbf{Trans.} & \textbf{LSTM} & \textbf{GRU} \\
    \hline
    \textbf{Experience Replay [LARS]}
      & $-41.824$ & $-41.927$ & $-41.737$
      & $-40.851$ & $-40.973$ & $-40.719$
      & $-38.910$ & $-39.004$ & $-38.890$ \\

    \textbf{Experience Replay [Reservoir]}
      & $-41.00$ & $-41.004$ & $-40.900$
      & $-38.750$ & $-38.954$ & $-38.700$
      & $-37.830$ & $-37.885$ & $-37.790$ \\

    \textbf{Loss Regularization [SI]}
      & $-41.003$ & $-41.042$ & $-40.850$
      & $-40.730$ & $-40.834$ & $-40.530$
      & $-39.650$ & $-39.731$ & $-39.530$ \\

    \textbf{Loss Regularization [EWC]}
      & $-39.220$ & $-39.271$ & $-39.150$
      & $-38.730$ & $-38.842$ & $-38.510$
      & $-37.610$ & $-37.835$ & $-37.550$ \\

    \textbf{Learning Without Forgetting}
      & $-35.900$ & $-36.500$ & $-35.800$
      & $-35.750$ & $-35.847$ & $-35.560$
      & $-34.500$ & $-34.673$ & $-34.420$ \\
    \hline
  \end{tabular}
  }
\end{table*}

\subsection{Learning without Forgetting (LwF)} \label{sec:lwf}
LwF offers a memory-free continual learning approach by distilling knowledge from a frozen copy of the model trained on past environments. After convergence on environment $T_1$ with dataset $\mathcal{D}_1$, the optimized parameters $\bm{\theta}$ are cloned as $\bm{\theta}_{\mathrm{old}} = \bm{\theta}$ and kept frozen to serve as a fixed ``teacher'' for subsequent tasks. For each subsequent environment \(T_k\) with a dataset \(\mathcal{D}_k\), as the UE hands over into each new UMi layout, whether moving from a compact to a dense or into a standard environment, we train the current model \(\bm{\theta}\) on mini-batches
\(\mathcal{B}_k \subset \mathcal{D}_k\)
to minimize a hybrid loss that combines the NMSE between ground-truth channels and the current model's predictions $\mathcal{L}_{\text{task}}$ with a distillation NMSE $\mathcal{L}_{\text{distill}}$ that measures the discrepancy between the current model's outputs and those of the frozen teacher.

\textbf{Mini-batch loss terms.}  
Let each sample in \(\mathcal{B}_k\) be
\((\mathbf{X}_i,\mathbf{H}_i)\), the standard NMSE between the current model's and the frozen teacher's predictions on the same inputs will be
\[
  \mathcal{L}_{\text{distill}}\bigl(\bm{\theta},\bm{\theta}_{\mathrm{old}},\mathcal{B}_k\bigr)
  = \frac{1}{|\mathcal{B}_k|}\sum_{i\in\mathcal{B}_k}
    \frac{\|\hat{\mathbf{H}}_i(\bm{\theta}_{\mathrm{old}}) - \hat{\mathbf{H}}_i(\bm{\theta})\|_F^2}{\|\hat{\mathbf{H}}_i(\bm{\theta_{\mathrm{old}}})\|_F^2}
\]
\textbf{Combined LwF Objective.}  
Algorithm~\ref{alg:lwf} reuses the mixing weight \(\lambda\in[0,1]\) from Experience Replay to balance fitting new channel data versus preserving past behavior:
\begin{equation}
\mathcal{L}_{\text{LwF}}(\bm{\theta})
= \;\lambda\;\mathcal{L}_{\text{task}}(\bm{\theta},\mathcal{B}_k)
\;+\;(1-\lambda)\;\mathcal{L}_{\text{distill}}(\bm{\theta},\bm{\theta}_{\mathrm{old}},\mathcal{B}_k).
\end{equation}
The hyperparameter $\lambda$ balances new‐cell adaptation and distillation: increasing $\lambda$ accelerates fitting to current fading and path‐loss, while decreasing $\lambda$ prioritizes alignment with the frozen teacher to preserve prior UMi‐cell characteristics. In this way, the model is encouraged to learn the fresh propagation characteristics of each new UMi scenario, capturing its unique path loss, multipath spread, and Doppler effects while still producing outputs that remain consistent to the behaviors learned in earlier layouts. This mechanism averts catastrophic forgetting across compact, dense, and standard scenarios while rapidly adapting to the UE’s evolving fading and path-loss statistics. ER replays the hardest fades through a loss-aware buffer; EWC supplements the $\mathrm{NMSE}$ loss with a Fisher-weighted penalty that anchors critical weights; LwF distills from a frozen teacher at zero memory cost. Collectively, these strategies span replay, regularization, and memory-free distillation, yielding complementary paths to handover-robust channel prediction.

% ER, EWC, and LwF each introduce a distinct continual-learning strategy tailored to channel prediction drift. ER leverages a loss-aware replay buffer that prioritizes rehearsing the hardest fades, directly interleaving raw past samples with new observations; EWC augments the NMSE loss with a Fisher-weighted quadratic penalty that anchors parameters critical to previously seen environments, penalizing only those weight changes likely to harm past performance; and LwF freezes a model snapshot as a teacher and adds a distillation-based NMSE term, preserving prior output behaviors without any extra memory footprint. Together, these methods span the trade-offs between explicit data replay, parameter regularization, and memory-free distillation, offering complementary paths to handover-robust wireless channel prediction.

\section{Results}
\label{Results}
\textbf{Experience Replay.} Figure~\ref{fig:experience_replay} compares two memory-based rehearsal schemes---uniform \emph{reservoir} sampling and \emph{loss-aware reservoir sampling} (LARS). 
Across the full SNR sweep (0-30\,dB) LARS consistently dominates the uniform buffer in all three propagation regimes. For the most challenging \textit{dense} case, LARS lowers the high-SNR NMSE floor from \(-39.7\)\,dB to \(-40.5\)\,dB, a \( \approx\! 20\% \) reduction in residual error, while the \textit{compact} setting benefits even more, reaching \(-42.1\)\,dB at 25 dB SNR. Because the buffer is biased towards hard-to-predict fades, the predictor rehearses precisely those outliers that dominate the tail of the loss distribution, yielding steeper convergence and a \(1\!-\!2\)\,dB gap over the uniform baseline throughout the mid-SNR region. These results confirm that targeted replay is crucial for retaining accurate channel dynamics when the user equipment (UE) traverses heterogeneous cell configurations.

\textbf{Loss Regularization Method.} Figure~\ref{fig:loss_regularization} evaluates Fisher-based EW against SI. While both mechanisms suppress catastrophic drift, SI achieves a uniformly lower NMSE, e.g.\ \(-40.8\)\,dB versus \(-39.3\)\,dB in the \textit{compact} layout at 20dB SNR. SI’s online importance tracking penalizes updates to weights with high past importance, enabling adaptation of peripheral parameters while preserving core ones. As a consequence, SI preserves the sharp error drop observed around 8-12dB without the saturation seen for EWC, yielding an additional \(0.8\!-\!1.4\)\,dB gain at high SNR and a \(10\%\) mean-squared-error reduction across the three tasks.

\textbf{Learning Without Forgetting.} Figure~\ref{fig:ewf} shows that \emph{LWF} delivers the \emph{highest} residual NMSE of the three continual-learning methods---e.g., \(-36.9\)\,dB (\textit{compact}), \(-35.8\)\,dB (\textit{dense}) and \(-34.4\)\,dB (\textit{standard}) at 25dB SNR---lagging LARS and SI by roughly \(0.7\!-\!1.5\)\,dB. Thus, even though its distillation-only strategy is less effective than replay or weight anchoring, LwF provides a lightweight improvement over no adaptation at all, requiring neither memory buffers nor second-order statistics and remaining attractive for resource-constrained deployments.

\section{Conclusion}
Our study demonstrates that continual learning substantially improves cross-cell channel prediction: loss-aware experience replay and synaptic intelligence lower the high-SNR NMSE floor by up to 2dB ($\approx$3\%) relative to naïve fine-tuning, while even memory-free LwF yields a consistent 1dB gain as shown in Table~\ref{tab:evaluation_losses}. These results confirm that rehearsing loss-critical fades and selectively anchoring influential weights are key to retaining past knowledge without sacrificing plasticity. The proposed framework therefore offers a practical path toward hand-over-robust CSI prediction, easing deployment across heterogeneous network topologies. For additional hyperparameter sensitivity studies, including sequence length effects and buffer-size ablations, see Appendix \ref{app:hyper_sensi} (Tables \ref{tab:evaluation_losses_16} and \ref{tab:er_memory_reduced}). Future work will focus on integrating the replay schedulers into 3GPP NR Channel State Information-Reference Signal (CSI-RS) procedures, quantifying fronthaul and UE memory overhead, and standardizing task-aware buffers within the O-RAN RIC for online adaptation. The approach scales to multi-cell joint prediction and can be benchmarked on upcoming 3GPP Release-19 indoor-hotspot and sub-6 GHz/sub-THz channel models, advancing standard-compliant, continually learning base-station intelligence.

% \section{Acknowledgements}
% This work is the result of a collaborative effort between Stanford University and industry leaders, including Intel Corporation, Ericsson, and Samsung Research America.

\bibliography{ref}
\bibliographystyle{icml2025}

% APPENDIX
\newpage
\appendix
\onecolumn
\section{Baseline Distributions and Results} \label{app:baseline_results}
\begin{figure*}[h!]
  \centering
  % ---------- first sub-figure ----------
  \begin{subfigure}[t]{0.44\textwidth}
      \centering
      \includegraphics[width=\textwidth]{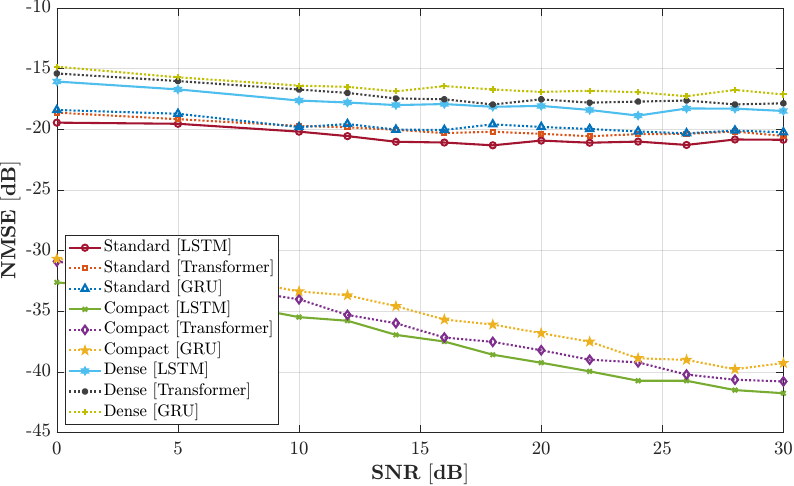}
      \caption{Baseline SNR trained on UMi compact and tested on all scenarios under all architectures.}
      \label{fig:baseline_compact}
  \end{subfigure}%
  \hspace{0.5em}
  % ---------- second sub-figure ----------
  \begin{subfigure}[t]{0.44\textwidth}
      \centering
      \includegraphics[width=\textwidth]{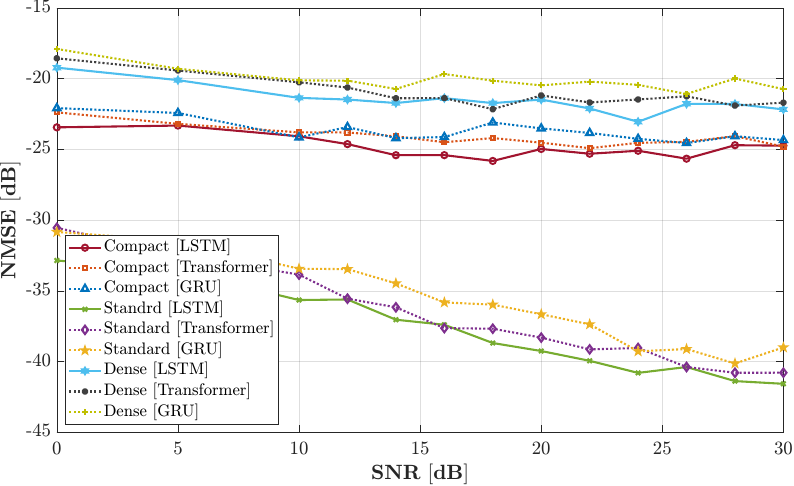}
      \caption{Baseline SNR trained on UMi standard and tested on all scenarios under all architectures.}
  \end{subfigure}%
  \hspace{0.5em}
  \caption{Prediction error under baseline conditions when tested under zero shot data settings.}
  \label{fig:baseline_standard}
\end{figure*}

First, discussion on the baseline models' architectures as discussed in Section~\ref{models}. Finding here elaborates on the architectures for the GRU and the Transformer and then discusses their performance in Figure~\ref{fig:baseline_standard}'s baseline scenarios.

\textbf{GRU.} The GRU variant retains the same preprocessing pipeline as the LSTM---each $T$-length window $\mathbf{X}\!\in\!\mathbb{C}^{2\times T\times N_{\mathrm{tx}}\times N_{\mathrm{rb}}\times N_{\mathrm{rx}}}$ is reshaped into a length-$d_{\text{in}}{\,=\,}2N_{\mathrm{tx}}N_{\mathrm{rb}}N_{\mathrm{rx}}$ feature vector per time step---but the temporal backbone is a three-layer Gated Recurrent Unit with hidden width $d_{\text{hid}}=32$. Compared with the LSTM, the GRU merges the input and forget gates, cutting the recurrent parameter count roughly by one-third while still maintaining gating dynamics that model slow and fast fading jointly. After processing, the final hidden state $\mathbf{h}_{T}\!\in\!\mathbb{R}^{d_{\text{hid}}}$ is mapped through a linear layer of size $(2N_{\mathrm{tx}}N_{\mathrm{rb}}N_{\mathrm{rx}})\!\times\!d_{\text{hid}}$ and reshaped back to $\hat{\mathbf{H}}\!\in\!\mathbb{C}^{2\times N_{\mathrm{tx}}\times N_{\mathrm{rb}}\times N_{\mathrm{rx}}}$. This leaner gating structure delivers competitive NMSE with shorter inference latency, making it attractive for edge deployment when computational budgets are tight.

\textbf{Transformer.} For longer temporal horizons, evaluation includes a lightweight Transformer that first flattens each spatial slice into a $d_{\text{in}}=2N_{\mathrm{tx}}N_{\mathrm{rb}}N_{\mathrm{rx}}$ vector, projects it to a $d_{\text{model}}=128$ embedding, and enriches it with a multi-frequency positional encoding tailored to wireless spectra. A single encoder layer and a single decoder layer, each with four self-attention heads, form the core sequence-to-sequence module; a learned start token serves as the one-step decoder query. The decoder output is passed through a final linear layer back to $2N_{\mathrm{tx}}N_{\mathrm{rb}}N_{\mathrm{rx}}$ and reshaped to the predicted channel matrix $\hat{\mathbf{H}}$. The self-attention mechanism allows the model to capture dependencies across the entire $T$-slot context without recurrence, enabling highly parallel training; however, its larger projection matrices and quadratic attention cost demand more memory than the recurrent baselines, so results limit depth to one layer each to stay within gNB resource constraints while still harnessing the Transformer's global receptive field.

\textbf{LSTM.} A \(3\)-layer LSTM network takes as input a tensor \(\mathbf{X}\!\in\!\mathbb{C}^{2\times T\times N_{\mathrm{tx}}\times N_{\mathrm{rb}}\times N_{\mathrm{rx}}}\), where the first dimension separates real and imaginary parts and \(T\) is the look-back window. At each time step, the spatial slice \(\bigl(2,N_{\mathrm{tx}},N_{\mathrm{rb}},N_{\mathrm{rx}}\bigr)\) flattens to a vector of length \(2N_{\mathrm{tx}}N_{\mathrm{rb}}N_{\mathrm{rx}}\), yielding a sequence whose feature size we denote \(d_{\text{in}}\). This sequence is processed by a stack of \(n_{\text{layers}}=3\) LSTM blocks with hidden width \(d_{\text{hid}}=32\), producing hidden states \(\mathbf{h}_{t}\!\in\!\mathbb{R}^{d_{\text{hid}}}\) for \(t=1,\ldots,T\). The final state \(\mathbf{h}_{T}\) is passed through a fully connected layer \(W\!\in\!\mathbb{R}^{(2N_{\mathrm{tx}}N_{\mathrm{rb}}N_{\mathrm{rx}})\times d_{\text{hid}}}\) to predict the next-slot channel matrix \(\hat{\mathbf{H}}\!\in\!\mathbb{C}^{2\times N_{\mathrm{tx}}\times N_{\mathrm{rb}}\times N_{\mathrm{rx}}}\), after which the original tensor shape is restored. The modest hidden width keeps the parameter count low---about \(4d_{\text{hid}}\bigl(d_{\text{in}}+d_{\text{hid}}\bigr)\) per layer---so the model fits comfortably in gNB memory while still capturing the essential channel dynamics. LSTMs excel in this setting because wireless channels form a strongly time-correlated sequence governed by user mobility and multipath evolution; the forget, input, and output gates allow the network to preserve long-range dependencies (slow fading) while rapidly adapting to short-term variations (fast fading) within the same recurrent architecture.

\section{Network Distributions.}
\label{distributions}
\subsection{Urban Macro Channel [UMa]}
For the UMa data set is synthesised with QuaDRiGa at \(f_c=2.6\;\mathrm{GHz}\) and \(20\;\mathrm{MHz}\) bandwidth, emulating macro base stations mounted \(25\;\mathrm{m}\) above street level that illuminate users at radial distances between \(100\) and \(500\;\mathrm{m}\). Three antenna-array configurations---\emph{standard} (\(8{\times}4\) panel, dual-pol dipoles), \emph{large\textendash H/V} (\(10{\times}6\) panel, quad-pol patches), and \emph{small\textendash V} (\(6{\times}2\) compact omni array)---capture a range of sector-capacity trade-offs, as summarised in Table~\ref{Uma}. Each Monte-Carlo realisation deploys \(256\) UEs whose linear tracks are sampled at \(30\) time instants and \(18\) OFDM resource blocks, yielding tensors of dimension \([30\times N_{\mathrm{tx}}\times 18\times N_{\mathrm{rx}}\times 256]\). Although the simulation seed is fixed to ensure reproducibility, every iteration perturbs the UE starting positions, and the LOS/NLOS state is drawn from the 3GPP distance-dependent law \(\Pr_{\mathrm{LOS}}(d)=e^{-\,d/300}\). This controlled displacement makes successive channel snapshots partially correlated, challenging the prediction network to learn both slow macro-scale trends and fast small-scale fading across the diverse UMa configurations.

\subsection{Urban Micro Channel [UMi]}
\begin{figure*}[t!]
  \centering
  % ---------- first sub-figure ----------
  \begin{subfigure}[t]{0.44\textwidth}
      \centering
      \includegraphics[width=\textwidth]{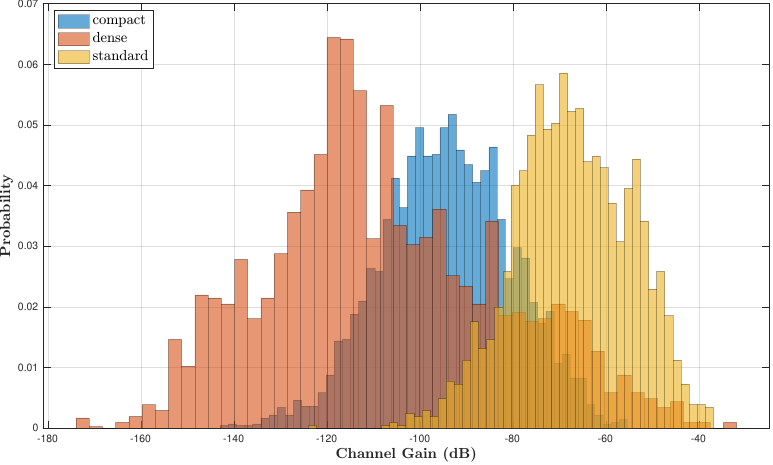}
      \caption{Probability distribution for UMi with paramters described in table~\ref{Umi}(a)}
  \end{subfigure}%
  \hspace{0.5em}
  % ---------- second sub-figure ----------
  \begin{subfigure}[t]{0.44\textwidth}
      \centering
      \includegraphics[width=\textwidth]{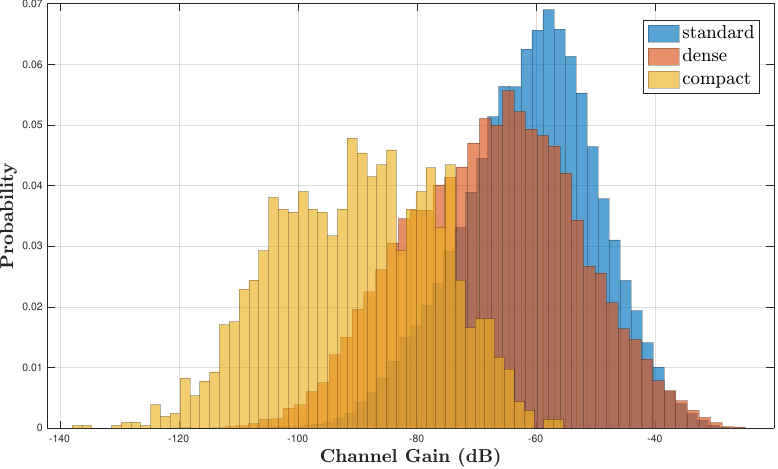}
      \caption{Probability distribution for UMa with paramters described in table~\ref{Uma}(b).}
  \end{subfigure}%
  \hspace{0.5em}
  \caption{Probability distribution deviation under different network configurations.}
  \label{fig:net_config_umi_uma}
\end{figure*}

The UMi data set used throughout the paper is generated with the QuaDRiGa Monte-Carlo engine configured for \(100\;\mathrm{MHz}\) bandwidth at \(f_c=5\;\mathrm{GHz}\) and an \(8\!\times\!2\) MIMO link (eight dual-polarised transmit elements arranged as a \(2{\times}2\) panel and a two-element UE array). Three propagation "flavors" are synthesized: \textit{standard}, \textit{dense}, and \textit{compact} as shown in Figure~\ref{fig:net_config_umi_uma}, which differ only in antenna downtilt, inter-element spacing, handset height, and the underlying 3GPP/5G channel profile (LOS or NLOS); the complete parameter list is given in Table~\ref{Umi}. Each simulation realizes \(256\) user equipment whose linear tracks are discretized into \(500\) time instants and \(18\) OFDM resource blocks, producing a complex-valued tensor of size \(\bigl[500\times 2\times 18\times 8\times 256\bigr]\). Although a fixed random seed guarantees repeatability, we introduce a small random displacement to every UE position at each Monte-Carlo iteration; this causes consecutive tensors to share local scatterers and therefore exhibit pronounced spatial-temporal correlation. The channel-prediction model must learn these correlations to extrapolate reliably from the recent \(T\)-slot history to the next-slot channel matrix, especially when the UE migrates between the three UMi scenarios.

\begin{table*}[t]
  \centering
  \caption{Configuration Parameters for 3GPP Urban Microcell (UMi) and Urban Macrocell (UMa) Scenarios}
  \label{tab:umi-uma-scenarios}

  %------------ left half : UMi ---------------------------------
  \begin{minipage}[b!]{0.48\textwidth}
    \centering
    \textbf{(a) UMi} \\[2pt]
    \label{Umi}
    \scalebox{0.8}{
    \begin{tabular}{lccc}
      \toprule
      \textbf{Parameter} & \textbf{Standard} & \textbf{Dense} & \textbf{Compact} \\
      \midrule
      Carrier Frequency            & 5\,GHz  & 5\,GHz & 5\,GHz \\
      Bandwidth                    & 100\,MHz & 100\,MHz & 100\,MHz \\
      Antenna Tilt (°)             & 30 & 10 & 0 \\
      Element Spacing ($\lambda$)  & 0.50 & 0.25 & 1.00 \\
      Rx Antenna Type              & dipole & patch & cross\_pol \\
      Rx Polarization              & $\pm45^\circ$ & H/V & $\pm45^\circ$ \\
      Distance Range (m)           & [50, 100] & [20, 60] & [120, 200] \\
      Tx Height (m)                & 10 & 6 & 15 \\
      UE Height (m)                & 1.5 & 1.0 & 2.0 \\
      \bottomrule
    \end{tabular}}
  \end{minipage}
  \hfill
  %------------ right half : UMa --------------------------------
  \begin{minipage}[b!]{0.48\textwidth}
    \centering
    \textbf{(b) UMa} \\[2pt]
    \label{Uma}
    \scalebox{0.8}{
    \begin{tabular}{lccc}
      \toprule
      \textbf{Parameter} & \textbf{Standard} & \textbf{Large--H/V} & \textbf{Small--V} \\
      \midrule
      Carrier Frequency            & 2.6\,GHz & 2.6\,GHz & 2.6\,GHz \\
      Antenna Tilt (°)             & 12 & 10 & 15 \\
      Element Spacing ($\lambda$)  & 0.50 & 0.60 & 0.50 \\
      Tx Array Size (M$\times$N)   & $8\times4$ & $10\times6$ & $6\times2$ \\
      Rx Antenna Type              & dipole & patch & omni \\
      Rx Polarization              & $\pm45^\circ$ & H/V & V \\
      Distance Range (m)           & [100, 500] & [100, 500] & [100, 500] \\
      Tx Height (m)                & 25 & 25 & 25 \\
      UE Height (m)                & 1.5 & 1.5 & 1.5 \\
      \bottomrule
    \end{tabular}}
  \end{minipage}
\end{table*}

\subsection{Dynamic Parametric Changes}

\begin{figure*}[b!]
  \centering
  % ---------- first sub-figure ----------
  \begin{subfigure}[t]{0.32\textwidth}
      \centering
      \includegraphics[width=\textwidth]{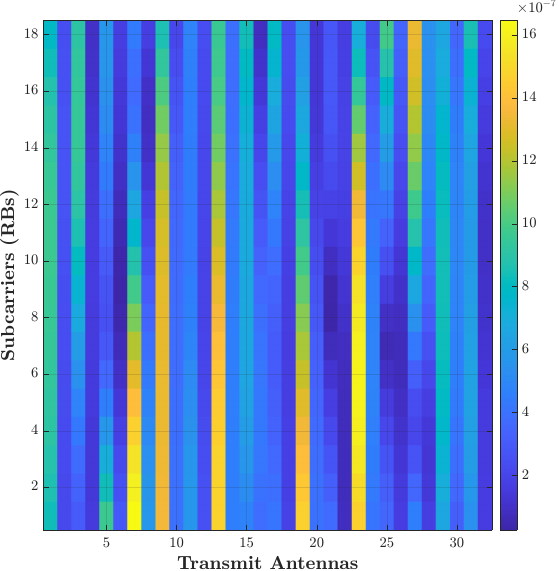}
      \caption{Transmit antenna vs Subcarriers for magntide of the complex channel matrix \textbf{H}.}
      \label{fig:subcarrier_magntiude}
  \end{subfigure}%
  \hspace{0.5em}
  % ---------- second sub-figure ----------
  \begin{subfigure}[t]{0.31\textwidth}
      \centering
      \includegraphics[width=\textwidth]{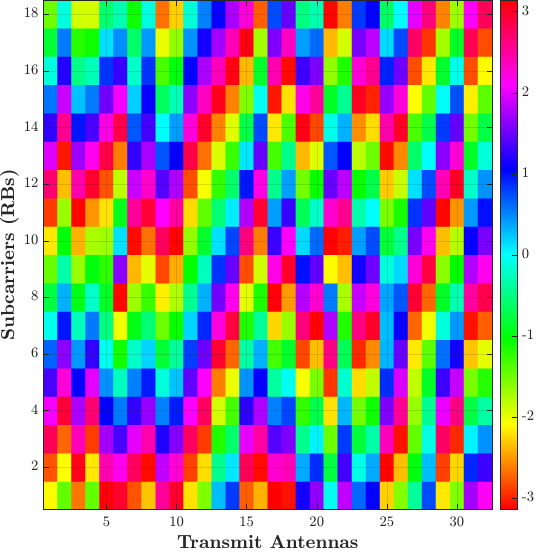}
      \caption{Transmit antenna vs Subcarriers for phase of the complex channel matrix \textbf{H}.}
      \label{fig:subcarrier_phase}
  \end{subfigure}%
  \hspace{0.5em}
  \begin{subfigure}[t]{0.32\textwidth}
      \centering
      \includegraphics[width=\textwidth]{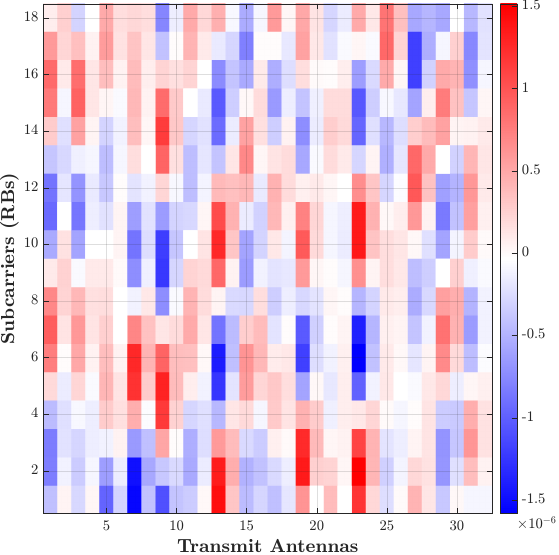}
      \caption{Transmit antenna vs Subcarriers for real part of the complex channel matrix \textbf{H}.}
      \label{fig:subcarrier_real_part}
  \end{subfigure}%
  \hspace{0.5em}
  \caption{Correlation among transmit anetnna vs. subcarriers for user 1 and time stamp 1.}
  \label{fig:correlation}
\end{figure*}

\textbf{1. Correlation among time series sequence.} For each UE the script draws a random radius 
$d\!\sim\!\mathcal U[d_{\min},d_{\max}]$  
and azimuth $\phi\!\sim\!\mathcal U[0,2\pi)$, fixes the initial position
\[
\mathbf p_{0}
     =\bigl[d\cos\phi,\;d\sin\phi,\;h_{\text{ue}}\bigr]^{\!\top},
\]
and then assigns the linear QuaDRiGa track
\verb|qd_track('linear',L,phi)| with length $L=2\,$m. 
The call 
\verb|tr.interpolate_positions(num_snap-1,'time')|  
creates equally spaced snapshots
\[
\mathbf p_{t}
     =\mathbf p_{0}
      +\frac{t}{T-1}\,L\,[\cos\phi,\;\sin\phi,\,0]^{\!\top},
      \quad t=0,\ldots,T-1,\;T=500,
\]
so the spatial step is 
$\Delta s=L/(T-1)\!\approx\!4\,$mm $\;(=\!{\sim}\lambda/15$ at $f_c=5\,$GHz). 
Under the WSSUS assumption an isotropic ring of scatterers of radius $R_{\rm sc}$ produces the small-scale correlation as shown in Figure~\ref{fig:correlation}.
\[
\rho_{\!h}(\Delta\mathbf r)
      =\mathbb E\!\bigl[h(\mathbf r)\,h^{\!*}(\mathbf r+\Delta\mathbf r)\bigr]
      =J_0\!\Bigl(\frac{2\pi}{\lambda}\,\|\Delta\mathbf r\|\Bigr),
\]
so between successive snapshots
$\rho_{\!h}(\Delta s)\!\approx\!J_0(2\pi/15)\!\gtrsim\!0.97$---\emph{very} strong. 
Because the UE displacement is deterministic, the channel at time~$t$
\[
H_t(k,m)=\sum_{\ell=1}^{L_{\text{cl}}}\!\alpha_\ell
  e^{-j2\pi k\Delta f\,\tau_{\ell}}
  e^{+j\frac{2\pi}{\lambda}d_t m\sin\theta_{\ell,t}}
  e^{-j\frac{2\pi}{\lambda}\,\mathbf k_\ell\!\cdot\!\mathbf p_t}
\]
inherits this coherence: the last term adds a linear phase
${\scriptstyle\propto}\,(\mathbf k_\ell\!\cdot\!\mathbf v)\,t$ whose slope is
so gentle that adjacent Monte-Carlo realisations remain \emph{temporally
self-correlated}. Consequently

\begin{itemize}[leftmargin=*]
\item the magnitude heat-map of Figure~\ref{fig:subcarrier_magntiude} shows
nearly vertical stripes---frequency coherence persists because
$\tau_{\ell,t}$ changes negligibly over $\Delta s$;
\item the phase map Figure~\ref{fig:subcarrier_phase} exhibits diagonal
ramps whose gradient equals the deterministic spatial phase drift
$\tfrac{2\pi}{\lambda}d_t\sin\theta_{\ell,t}$ multiplied by the constant
step $\Delta s$;
\item the real part in Figure~\ref{fig:subcarrier_real_part} reveals blocks of coherent
(additive) or destructive interference depending on whether the phase
difference between neighbouring antennas is near $0$ or~$\pi$.
\end{itemize}

In sum, the \emph{incremental UE motion}
$\mathbf p_{t+1}\!-\!\mathbf p_t=\Delta s\,[\cos\phi,\sin\phi,0]^{\!\top}$
couples the spatial, frequency and \emph{time} dimensions: adjacent
snapshots share scatterers and remain highly correlated, providing the
predictive model with exploitable temporal structure.

\begin{figure*}[t!]
  \centering
  % ---------- first sub-figure ----------
  \begin{subfigure}[t]{0.32\textwidth}
      \centering
      \includegraphics[width=\textwidth]{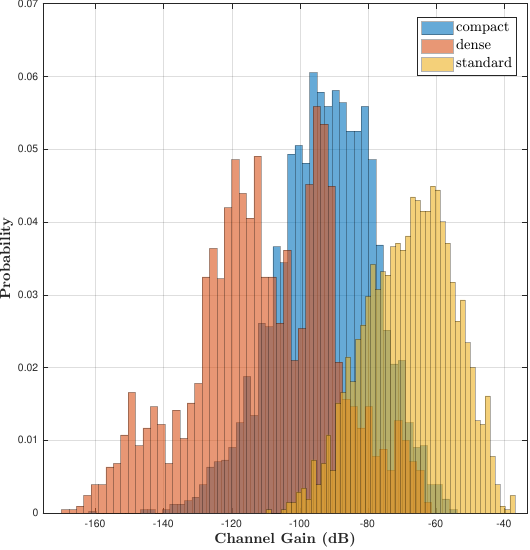}
      \caption{Probability distribution for UMi with element spacing [$\lambda$ = 1, 2, 5].}
  \end{subfigure}%
  \hspace{0.5em}
  % ---------- second sub-figure ----------
  \begin{subfigure}[t]{0.32\textwidth}
      \centering
      \includegraphics[width=\textwidth]{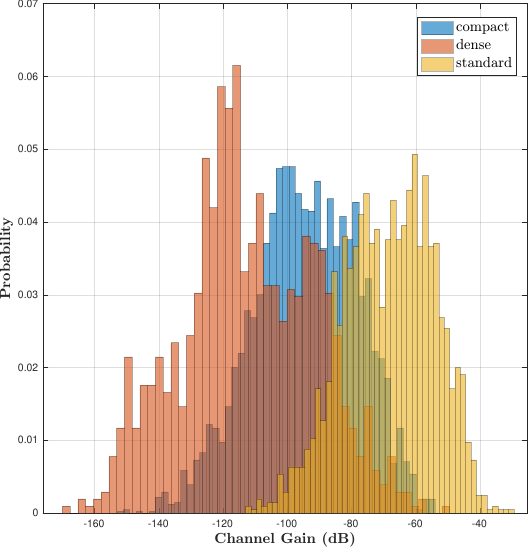}
      \caption{Probability distribution for UMi with element spacing [$\lambda$ = 0.5 , 0.025, 3.0]}
  \end{subfigure}%
  \hspace{0.5em}
  \begin{subfigure}[t]{0.32\textwidth}
      \centering
      \includegraphics[width=\textwidth]{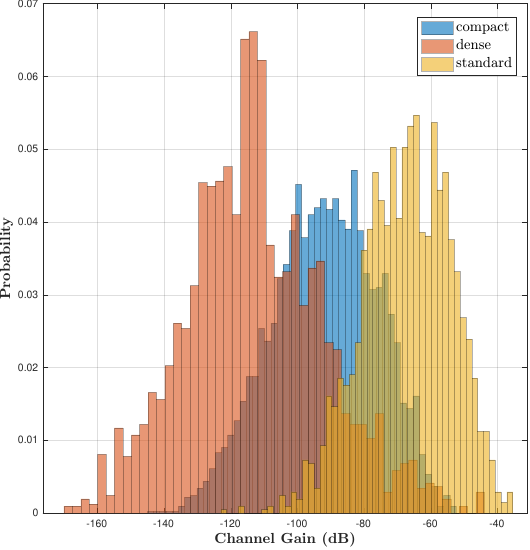}
      \caption{Probability distribution for UMi with element spacing [$\lambda$ = 3.5, 1, 4.2]}
  \end{subfigure}%
  \hspace{0.5em}

    % ---------- first sub-figure ----------
  \begin{subfigure}[t]{0.32\textwidth}
      \centering
      \includegraphics[width=\textwidth]{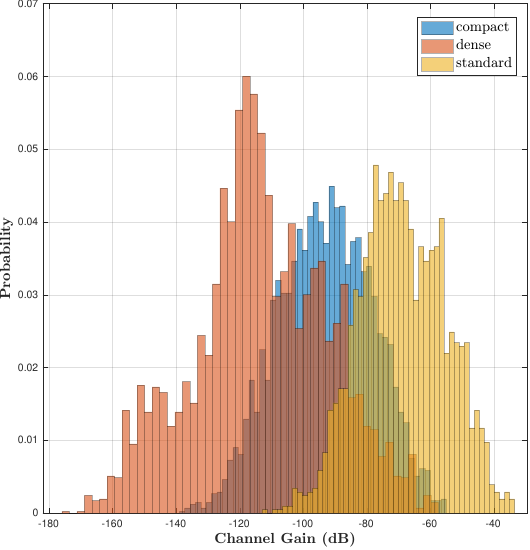}
      \caption{Probability distribution for UMi with antenna tilt [ ° = 0, 90, 120].}
      \label{fig:antenna_tilt_1}
  \end{subfigure}%
  \hspace{0.5em}
  \begin{subfigure}[t]{0.32\textwidth}
      \centering
      \includegraphics[width=\textwidth]{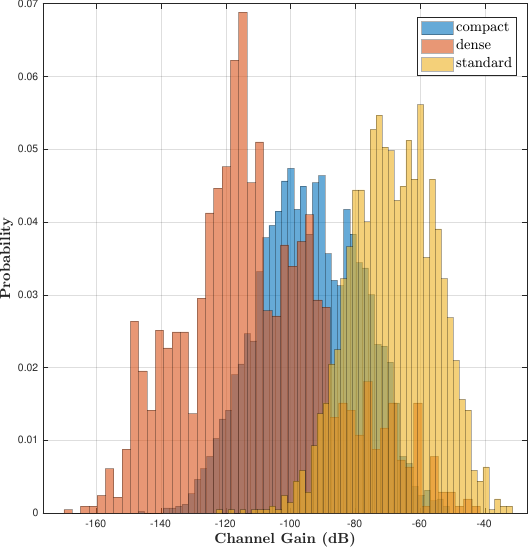}
      \caption{Probability distribution for UMi with antenna tilt [° = 0, 10 , 15]}
      \label{fig:antenna_tilt_2}
  \end{subfigure}%
  \hspace{0.5em}
  \begin{subfigure}[t]{0.32\textwidth}
      \centering
      \includegraphics[width=\textwidth]{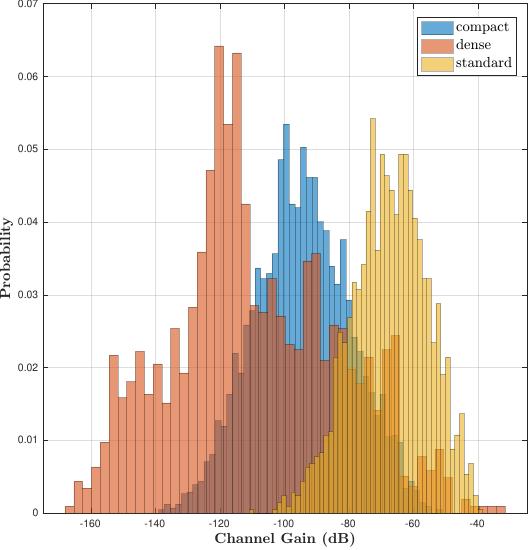}
      \caption{Probability distribution for UMi with antenna tile [° = 0, 100, 200]}
      \label{fig:antenna_tilt_3}
  \end{subfigure}%

  \caption{Effect of element spacing and antenna tilt on probability distribution of channel gains under UMi scenario.}
  \label{fig:perf-comparison_elementz}
\end{figure*}

\textbf{2. Effect of Element Spacing.} Increasing the inter-element spacing \(d\lambda\) widens the physical aperture of the array---yielding an \(L/\lambda\)-dependent array gain that shifts the entire channel gain histogram to the right---while simultaneously driving down the spatial correlation terms \(\rho_k(d)=J_0(2\pi d k)\) in the transmit/receive correlation matrix. Because the variance of the instantaneous power \(g=\|\mathbf R^{1/2}(d)\mathbf w\|^{2}\) is \(2\sum_{k}(N-k)\rho_k^{2}(d)\), weaker correlations at larger \(d\) shrink this variance, producing a tighter distribution with fewer deep fades. Hence, ``dense'' arrays with \(d\!\lesssim\!0.5\lambda\) exhibit the left-shifted, broad brown histograms observed, ``compact'' (\(d\!\approx\!1\lambda\)) fall in the middle, and ``standard'' spacings (\(d\!\ge\!2\lambda\)) give the right-shifted, sharply peaked yellow curves as shown in Figure~\ref{fig:perf-comparison_elementz}.

\textbf{3. Effect of Antenna Tilt.} Downtilting the base-station panel multiplies every small-scale channel coefficient by the vertical antenna pattern
\[
G(\theta)=G_{\max}-\min\!\Bigl\{\,12\!\bigl(\tfrac{\theta-\theta_{\text{tilt}}}{\theta_{3\mathrm{dB}}}\bigr)^{2},\;{\rm SLA}_{V}\Bigr\}\;[\mathrm{dB}],
\]
with half-power beam-width $\theta_{3\mathrm{dB}}\!\approx\!10^{\circ}$ and a side-lobe cap ${\rm SLA}_{V}\!\approx\!20\,$dB. 
For a UE at horizontal range $R$ and height $h_{\mathrm{ue}}$ below the array ($h_{\mathrm{tx}}$), the elevation angle is $\theta(R)=\tan^{-1}\!\bigl[(h_{\mathrm{tx}}-h_{\mathrm{ue}})/R\bigr]$, so the large-scale channel gain
\[
g_{\mathrm{LS}}(R)=
G\!\bigl(\theta(R)\bigr)-10\,\alpha\log_{10}R+X_{\sigma}\quad[\mathrm{dB}],
\]
combines tilt-dependent antenna gain, distance-dependent path-loss ($\alpha\!\approx\!3.1$ in UMi NLOS) and log-normal shadowing $X_{\sigma}\sim\mathcal N(0,\sigma^{2})$. 
With UEs roughly uniform in range $[R_{\min},R_{\max}]$, the elevation pdf is 
\[
f_{\Theta}(\theta)=\frac{(h_{\mathrm{tx}}-h_{\mathrm{ue}})\cos^{2}\theta}{(R_{\max}-R_{\min})(h_{\mathrm{tx}}-h_{\mathrm{ue}})\cos\theta+R_{\min}},
\]
so the variance $\operatorname{Var}\bigl[G(\Theta)\bigr]=\mathbb E[G^{2}(\Theta)]-\mathbb E^{2}[G(\Theta)]$ decreases as $\theta_{\text{tilt}}$ increases because the effective support of $\Theta$ shrinks to the nearly flat main-lobe. Hence large downtilt (\(\sim\!30^{\circ}\)) aligns the main lobe with nearby UEs, producing higher median gains and a tighter histogram (yellow ``standard'' curves); little or no downtilt (\(0\!-\!10^{\circ}\)) illuminates farther ranges, yielding lower, more spread-out gains (brown ``dense'' curves); intermediate tilts (blue ``compact'') lie between these extremes. The curves are shown in Figure~\ref{fig:perf-comparison_elementz}.

\begin{figure*}[t!]
  \centering
  % ---------- first sub-figure ----------
  \begin{subfigure}[t]{0.32\textwidth}
      \centering
      \includegraphics[width=\textwidth]{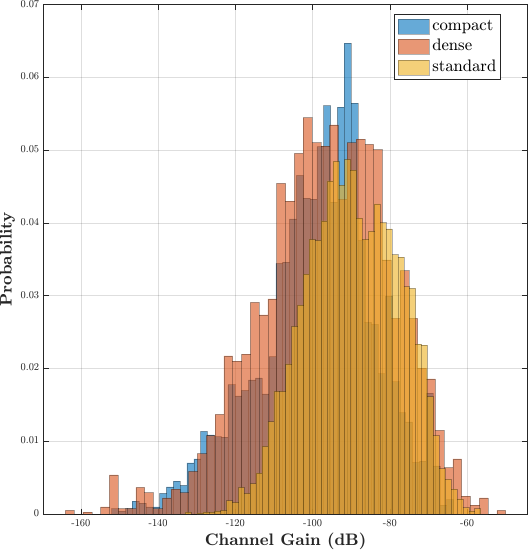}
      \caption{Channel gain vs. Probability distribution for scenario A.}
  \end{subfigure}%
  \hspace{0.5em}
  % ---------- second sub-figure ----------
  \begin{subfigure}[t]{0.32\textwidth}
      \centering
      \includegraphics[width=\textwidth]{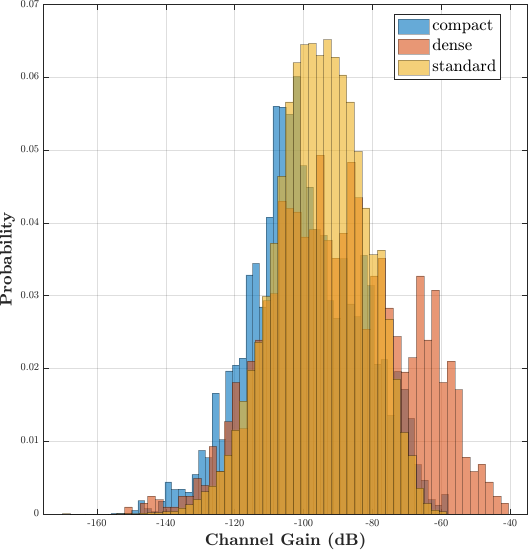}
      \caption{Channel gain vs. Probability distribution for scenario B.}
  \end{subfigure}%
  \hspace{0.5em}
  \begin{subfigure}[t]{0.32\textwidth}
      \centering
      \includegraphics[width=\textwidth]{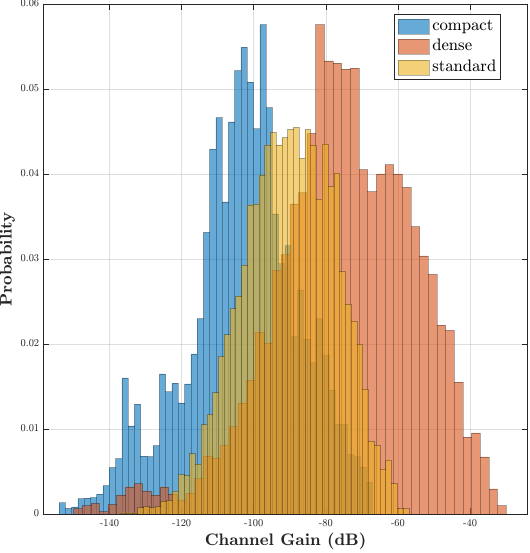}
      \caption{Channel gain vs. Probability distribution for scenario C.}
  \end{subfigure}%
  \hspace{0.5em}
  \caption{Combined effect of cross network configurtion on channel distributions.}
  \label{fig:cross_network_config}
\end{figure*}

\textbf{4. Combined Effect of Cross Parameterization.} Experiments examine three representative environments whose antenna deployments differ widely in aperture, element spacing, height, downtilt, and user range.

\textbf{Scenario A.}  
This baseline group follows the canonical 3GPP-UMi geometry. The \emph{standard\_A} cell employs an $8\times2$ dual-polarised panel ($MN=16$ elements) mounted at $h_{\mathrm{tx}}=35\,$m, downtilted by $6^{\circ}$ and spaced at $d_t=\tfrac23\lambda$; UEs move at ranges $R\in[80,150]\,$m under the LOS profile \textit{3GPP\_38.901\_UMi\_LOS}. 
The \emph{dense\_A} flavour shrinks the array to a $4\times1$ ($MN=4$) patch with no downtilt and very tight $0.25\lambda$ spacing, placed at $h_{\mathrm{tx}}=10\,$m; pedestrians roam only $R\in[10,60]\,$m in NLOS (\textit{3GPP\_38.901\_UMi\_NLOS}). 
Finally, the \emph{compact\_A} cell restores an $8\times4$ panel ($MN=32$) but widens the spacing to $2.5\lambda$ and downtilts by $30^{\circ}$; at $h_{\mathrm{tx}}=25\,$m it serves users at $R\in[40,100]\,$m with the LOS-rich \textit{5G-ALLSTAR\_DenseUrban\_LOS} profile.

\textbf{Scenario B.}  
This setting exaggerates macro/ hot-spot contrast. 
The \emph{standard\_B} macro site is a $16\times4$ array ($MN=64$) on a $45\,$m rooftop, downtilt $8^{\circ}$, wide $1.2\lambda$ spacing, covering $R\in[150,300]\,$m in LOS. 
Opposite to it, the \emph{dense\_B} pedestrian hot-spot uses just $2\times1$ elements ($MN=2$) at street-lamp height $h_{\mathrm{tx}}=6\,$m, zero tilt, ultra-tight $0.15\lambda$ spacing and NLOS users at $R\in[5,30]\,$m. 
Mid-way, the \emph{compact\_B} cell keeps an $8\times4$ panel ($MN=32$) with \emph{huge} $3\lambda$ spacing and strong $35^{\circ}$ downtilt, serving $R\in[40,120]\,$m under LOS.

\textbf{Scenario C.}  
A suburban/indoor mix spanning indoor hotspots to rooftop cells: \emph{standard\_C}: $8\times2$ array ($MN=16$) at $h_{\mathrm{tx}}=25\,$m, no downtilt, spacing $1\lambda$, LOS users $R\in[100,180]\,$m. \emph{dense\_C}: indoor $4\times2$ panel ($MN=8$) at $h_{\mathrm{tx}}=3\,$m, upward tilt $-15^\circ$, spacing $0.5\lambda$, NLOS users $R\in[2,15]\,$m. \emph{compact\_C}: rooftop $12\times3$ array ($MN=36$) at $h_{\mathrm{tx}}=30\,$m, downtilt $20^\circ$, spacing $2\lambda$, LOS users $R\in[60,140]\,$m.

The wide variation in array aperture, element spacing, antenna height, downtilt angle, and user-to-base-station distance across Scenarios A-C produces fundamentally different channel ``shapes'' in each environment as shown in Figure~\ref{fig:cross_network_config}. For instance, a dense urban hotspot (e.g., \emph{dense\_C}) uses tightly spaced indoor antennas and very short links, resulting in richly scattered, rapidly varying multipath that is hard to predict. By contrast, a rooftop macrocell (e.g., \emph{standard\_C}) with wider spacing, higher mounting, and longer ranges yields smoother, largely line-of-sight channels. Steep downtilts and greater heights enhance broad coverage ideal in suburban settings, while close spacing and low mounts capture fine multipath detail critical for indoor NLOS operation. Since our neural predictor internalizes the specific gain distribution, delay spread, and angular statistics of one scenario, it struggles when faced with a channel whose clutter density, pathloss characteristics, and angular spread differ significantly. In simple terms, changing physical parameters reshapes how signals bounce and fade; without exposure to each unique ``network configuration settings,'' model accuracy (NMSE) and temporal forecasting degrade significantly.

\section{Hyperparameter Sensitivity in Continual Learning} \label{app:hyper_sensi}

\subsection{Impact of Sequence Length on Continual Learning Pipelines}

Table~\ref{tab:evaluation_losses_16} compares the NMSE for the above-discussed five dynamic continual-learning pipelines with sequence length=16. In every case, the LSTM backbone outperforms the Transformer, which in turn outperforms the GRU. Performance degrades smoothly from the Compact to Dense to Standard environments. Moreover, both ER variants (LARS and Reservoir) consistently achieve the lowest NMSE, followed by EWC, with lwf exhibiting the highest errors.
  \begin{table*}[h!]
  \centering
  \caption{\textbf{Evaluation loss comparison under dynamic continual learning pipelines (sequence length = 16, replay memory size = 5000) [NMSE Loss in dB]}}
  \label{tab:evaluation_losses_16}
  \vspace{2pt}
  \scalebox{0.8}{
  \begin{tabular}{l|ccc|ccc|ccc}
    \hline
    \textbf{Continuous Learning Pipelines}
      & \multicolumn{3}{c|}{\textbf{Test: UMi Compact}}
      & \multicolumn{3}{c|}{\textbf{Test: UMi Dense}}
      & \multicolumn{3}{c}{\textbf{Test: UMi Standard}} \\
    \cline{2-10}
      & \textbf{Trans.} & \textbf{LSTM} & \textbf{GRU}
      & \textbf{Trans.} & \textbf{LSTM} & \textbf{GRU}
      & \textbf{Trans.} & \textbf{LSTM} & \textbf{GRU} \\
    \hline
    \textbf{Experience Replay [LARS]}
      & $-41.624$ & $-41.727$ & $-41.537$
      & $-40.651$ & $-40.773$ & $-40.519$
      & $-38.710$ & $-38.804$ & $-38.690$ \\

    \textbf{Experience Replay [Reservoir]}
      & $-40.800$ & $-40.804$ & $-40.700$
      & $-38.550$ & $-38.754$ & $-38.500$
      & $-37.630$ & $-37.685$ & $-37.590$ \\

    \textbf{Loss Regularization [SI]}
      & $-40.803$ & $-40.842$ & $-40.650$
      & $-40.530$ & $-40.634$ & $-40.330$
      & $-39.450$ & $-39.531$ & $-39.330$ \\

    \textbf{Loss Regularization [EWC]}
      & $-39.020$ & $-39.071$ & $-38.950$
      & $-38.530$ & $-38.642$ & $-38.310$
      & $-37.410$ & $-37.635$ & $-37.350$ \\

    \textbf{Learning Without Forgetting}
      & $-35.700$ & $-36.300$ & $-35.600$
      & $-35.550$ & $-35.647$ & $-35.360$
      & $-34.300$ & $-34.473$ & $-34.220$ \\
    \hline
  \end{tabular}}
\end{table*}

\subsection{Effect of Reduced Replay Buffer Size on ER Performance}
Table~\ref{tab:er_memory_reduced} reports the NMSE for ER with the memory buffer reduced from 5000 to 3000 and sequence $length=32$. As before, the LSTM backbone consistently outperforms the Transformer, which in turn beats the GRU, and performance degrades smoothly from the Compact to Dense to Standard UMi scenarios.
\begin{table*}[h!]
  \centering
  \setlength{\tabcolsep}{2pt}
  \caption{\textbf{Evaluation loss for Experience Replay pipelines with reduced memory (\(M=3000\)) and sequence length = 32 [NMSE Loss in dB]}}
  \label{tab:er_memory_reduced}
  \vspace{2pt}
  \scalebox{0.8}{
  \begin{tabular}{l|ccc|ccc|ccc}
    \hline
    \textbf{ER Variant}
      & \multicolumn{3}{c|}{\textbf{Test: UMi Compact}}
      & \multicolumn{3}{c|}{\textbf{Test: UMi Dense}}
      & \multicolumn{3}{c}{\textbf{Test: UMi Standard}} \\
    \cline{2-10}
      & \textbf{Trans.} & \textbf{LSTM} & \textbf{GRU}
      & \textbf{Trans.} & \textbf{LSTM} & \textbf{GRU}
      & \textbf{Trans.} & \textbf{LSTM} & \textbf{GRU} \\
    \hline
    \textbf{Experience Replay [LARS]}
      & $-40.317$ & $-41.289$ & $-39.137$
      & $-39.682$ & $-40.527$ & $-38.314$
      & $-38.201$ & $-38.864$ & $-36.873$ \\

    \textbf{Experience Replay [Reservoir]}
      & $-39.612$ & $-40.574$ & $-38.432$
      & $-37.976$ & $-38.815$ & $-36.752$
      & $-36.443$ & $-37.311$ & $-35.216$ \\
    \hline
  \end{tabular}}
\end{table*}

\end{document}